\documentclass[iop,twocolappendix]{emulateapj}
\usepackage{graphicx}
\usepackage{epstopdf}
\usepackage{appendix,natbib}
\usepackage[colorlinks=true,citecolor=blue,linkcolor=blue,urlcolor=blue]{hyperref} 

\def\degree{\ifmmode {^\circ}\else {$^\circ$}\fi}
\def\rstar{\ifmmode {\, R_{\star}}\else $R_{\star}$\fi}
\def\msol{\ifmmode {\, M_{\odot}}\else $\text{M}_{\odot}$\fi}
\def\rsol{\ifmmode {\, R_{\odot}}\else $R_{\odot}$\fi}
\def\lsol{\ifmmode {\, L_{\odot}}\else $L_{\odot}$\fi}
\def\msolyr{\ifmmode {\, M_{\odot}\,{\rm yr}^{-1}}\else $M_{\odot}\,{\rm yr}^{-1}$\fi}
\def\mdot{\ifmmode {\,\dot{M}}\else $\dot{M}$\fi}
\def\mdotyr{\ifmmode {\,\dot{M}\,yr^{-1}}\else $\dot{M}\,yr^{-1}$\fi}

\newcommand{\OIII}{[{\ion{O}{3}}]}

\newcommand{\Hb}{{H$\beta$}}

\newcommand{\NII}{[{\ion{N}{2}}]}
\newcommand{\NIII}{[{\ion{N}{2}}]~$\lambda$6585}
\newcommand{\NIIII}{[{\ion{N}{2}}]~$\lambda$6550}
\newcommand{\SII}{[{\ion{S}{2}}]}
\newcommand{\SIIl}{[\ion{S}{2}]~$\lambda \lambda$6718,6733}
\newcommand{\SIIratio}{[\ion{S}{2}]~$\lambda$6718/[\ion{S}{2}]~$\lambda$6733}

\def\O4363{[{\ion{O}{3}}]~$\lambda$4364}

\newcommand{\Ha}{{H$\alpha$}}

\begin{document}

\title{The MOSDEF Survey:  Broad Emission Lines at $\MakeLowercase{z}=1.4 - 3.8$*}
\author{William R. Freeman\altaffilmark{1}, Brian Siana\altaffilmark{1}, Mariska Kriek\altaffilmark{2}, Alice E. Shapley\altaffilmark{3}, Naveen Reddy\altaffilmark{1}, Alison L. Coil\altaffilmark{4}, Bahram Mobasher\altaffilmark{1}, Alexander L. Muratov\altaffilmark{4}, Mojegan Azadi\altaffilmark{4}, Gene Leung\altaffilmark{4}, Ryan Sanders\altaffilmark{3}, Irene Shivaei\altaffilmark{1,5}, Sedona H. Price\altaffilmark{2,6},  Laura DeGroot\altaffilmark{7}, Du\v{s}an Kere\v{s}\altaffilmark{4}}

\altaffiltext{}{email: billfreeman44@yahoo.com}
\altaffiltext{*}{Based on data obtained at the W.M. Keck Observatory, which is operated as a scientific partnership among the California Institute of Technology, the University of California, and NASA, and was made possible by the generous financial support of the W.M. Keck Foundation.}
\altaffiltext{1}{Department of Physics \& Astronomy, University of California, Riverside, 900 University Avenue, Riverside, CA 92521, USA}
\altaffiltext{2}{Astronomy Department, University of California, Berkeley, CA 94720, USA}
\altaffiltext{3}{Department of Physics \& Astronomy, University of California, Los Angeles, 430 Portola Plaza, Los Angeles, CA 90095, USA}
\altaffiltext{4}{Center for Astrophysics and Space Sciences, University of California, San Diego, 9500 Gilman Dr., La Jolla, CA 92093- 0424, USA}
\altaffiltext{5}{Steward Observatory, University of Arizona, Tucson, AZ 85721, USA}
\altaffiltext{6}{Max-Planck-Institut für extraterrestrische Physik, Giessenbachstr. 1,D-85737 Garching, Germany}
\altaffiltext{7}{Department of Physics, The College of Wooster, Wooster, OH 44691}

\begin{abstract}
We present results from the MOSFIRE Deep Evolution Field (MOSDEF) survey on broad flux from the nebular emission lines \Ha, \NII, \OIII, \Hb, and \SII.  The sample consists of 127 star-forming galaxies at   $1.37 < z < 2.61$ and 84 galaxies at  $2.95 < z < 3.80$.   We decompose the emission lines using narrow ($\text{FWHM} < 275 \ \text{km s}^{-1}$) and broad ($\text{FWHM} > 300 \ \text{km s}^{-1}$) Gaussian components for individual galaxies and stacks.      Broad emission  is detected at $>3\sigma$  in $<10$\% of galaxies and the broad flux accounts for 10-70\% of the total flux.  We find a slight increase in broad to narrow flux ratio with mass but note that we cannot reliably detect broad emission with $\text{FWHM} < 275 \ \text{km s}^{-1}$, which may be significant at low masses.   Notably, there is a correlation between higher signal-to-noise (S/N) spectra and a broad component detection indicating a S/N dependence in our ability to detect broad flux.
When placed on the N2-BPT diagram (\OIII/\Hb \ vs. \NII/\Ha)  the broad components of the stacks are shifted towards  higher \OIII/\Hb \ and \NII/\Ha \ ratios compared to the narrow component.  We compare the location of the broad components to shock models and find that the broad component could be explained as a shocked outflow, but we do not rule out other possibilities such as the presence of an AGN.
We estimate the mass loading factor (mass outflow rate/star formation rate) assuming the broad component is  a photoionized outflow and find that the mass loading factor increases  as a function of mass which agrees with previous studies.  
We show that adding emission from shocked gas to $z\sim0$ SDSS spectra shifts galaxies towards the location of $z\sim2$ galaxies on several emission line diagnostic diagrams.


\end{abstract}

\section{Introduction}
\setcounter{footnote}{0} 

Rest-frame optical nebular emission lines such as \Ha, \NII, \OIII, \Hb, and \SII \ are diagnostics of physical properties of galaxies such as star formation rate (SFR) (\citealt{1998ARA&A..36..189K}, \citealt{2015arXiv150703017S}), dust extinction (\citealt{2013ApJ...777L...8K}, \citealt{2014ApJ...795..165S}, \citealt{2015arXiv150402782R}), electron density (\citealt{1989agna.book.....O}, \citealt{2009ApJ...701...52H}, \citealt{2010ApJ...725.1877B}, \citealt{2016ApJ...816...23S}), and metallicity (\citealt{2004MNRAS.348L..59P}, \citealt{2006ApJ...644..813E}, \citealt{2015ApJ...799..138S}).   Consequently, galaxies fall into well-defined patterns in emission line diagnostic diagrams such as the N2-BPT diagram (\OIII/\Hb \ vs. \NII/\Ha) and the S2-BPT diagram (\OIII/\Hb \ vs. \SII/\Ha) (\citealt{1981PASP...93....5B}).  The position of a galaxy on these diagrams is determined by its underlying physical conditions such as electron density, hardness of ionizing radiation, AGN presence, and metallicity (\citealt{2013ApJ...774..100K}, \citealt{2015ApJ...801...88S}, \citealt{2014arXiv1409.6522C}, \citealt{2016ApJ...816...23S}).

Galaxies at $z\sim2$ show a systematic offset from local galaxies in the N2-BPT diagram (\citealt{2005ApJ...635.1006S}, \citealt{2006ApJ...644..813E}, \citealt{2008ApJ...678..758L}).  There has been a great deal of study on the cause of this offset.  Plausible explanations for the offset  include higher ionization parameters (\citealt{2008MNRAS.385..769B}, \citealt{2015ApJ...812L..20K}), elevated N/O ratios at fixed O/H (\citealt{2016ApJ...828...18M}, \citealt{2015ApJ...801...88S}, \citealt{2015ApJ...813..126J}, \citealt{2016ApJ...816...23S}, \citealt{2016ApJ...817...57C}), harder stellar radiation fields at fixed nebular metallicity (\citealt{2014ApJ...795..165S}, \citealt{2017ApJ...836..164S}), and different star formation histories (\citealt{2016ApJ...826..159S}, \citealt{2017arXiv170600010H}).

The dynamics of galaxies can be used to further understand the origin of the line emission and thus the location $z\sim2$ galaxies in the BPT diagram. Broad emission relative to the intrinsic galactic emission is of particular interest, as it is indicative of outflowing material (\citealt{2005ARA&A..43..769V}, \citealt{1990ApJS...74..833H}, \citealt{2010ApJ...713..686D}, \citealt{2011ApJ...742...11S}, \citealt{2013ApJ...768...74T}, \citealt{2009ApJ...701..955S}, \citealt{2012ApJ...761...43N}, \citealt{2014ApJ...781...21N}, \citealt{2014ApJ...787...38F}, \citealt{2011ApJ...733..101G}).  Broad emission is also seen in some AGNs that is not emitted from the broad line region, but from outflows (\citealt{2014ApJ...787...38F}, \citealt{2014ApJ...796....7G}, \citealt{2017arXiv170310255L}).  In the local universe, broad emission is seen in luminous infrared galaxies (\citealt{2005ApJS..160..115R}, \citealt{2012MNRAS.424..416W}).  Luminous infrared galaxies are typically associated with active or recent mergers and the increased star formation associated with merging could be driving the outflows.  Estimates of physical properties of galaxies from emission lines typically assume the emission originates in the HII regions of galaxies.  If a significant portion of the flux originates in a broad, outflowing component, this might influence the estimates of galaxy physical properties.

It is important to understand the effects of broad emission on measurements of physical properties of galaxies, particularly at $z\sim2$, where they have higher star formation rate (SFR) (\citealt{2009ApJ...692..778R}, \citealt{1998ApJ...498..106M}, \citealt{2014ARA&A..52..415M}), higher gas fractions (\citealt{2010ApJ...713..686D}, \citealt{2011ApJ...742...11S}, \citealt{2013ApJ...768...74T}), and are more compact (\citealt{2005ApJ...630L..17T};\citealt{2003MNRAS.343..978S}, \citealt{2005ApJ...635..959B}) compared to galaxies at $z\sim0$.  The higher SFR and smaller sizes lead to a larger star formation surface density ($\Sigma_{\text{SFR}}$) which may result in an outflow (\citealt{2011ApJ...731...41O}, \citealt{2012ApJ...761...43N}).   Studies of broad emission at $z\sim2$ have shown that broad emission is more prominent above a star formation surface density ($\Sigma_{\text{SFR}}$) of 1.0 \msol/yr/kpc$^2$ (\citealt{2012ApJ...761...43N}).    However, previous studies of broad emission at this redshift are based on a small sample (\citealt{2014ApJ...781...21N}), are done on galaxies with high sSFR (\citealt{2009ApJ...701..955S}, \citealt{2012ApJ...761...43N}), or focus on galaxies with AGN (\citealt{2014ApJ...787...38F}, \citealt{2011ApJ...733..101G}, \citealt{2017arXiv170310255L}). A study of broadened emission with a large sample of typical galaxies at $z\sim1-3$ is necessary to understand how broad emission affects the average star-forming galaxy.

We use the MOSFIRE Deep Evolution Field (MOSDEF\footnote{\url{http://mosdef.astro.berkeley.edu/}}) survey (\citealt{2015ApJS..218...15K}) to study broadened emission for a large sample of $z\sim1-3$ galaxies.  We  obtained near-infrared spectra for $\sim1500$ high-redshift galaxies using the MOSFIRE instrument (\citealt{2012SPIE.8446E..0JM}) on the W. M. Keck telescope.   These spectra enable measurements of the rest-frame optical nebular emission lines for galaxies at  $1.37<z<3.80$.  The data in the MOSDEF survey allow for measurements of broad emission on a large sample of star-forming galaxies (as well as AGN, presented in \citealt{2017arXiv170310255L}).  The goal of this paper is to measure or place limits on a broad component in star-forming galaxies in order to determine the amount of broad emission in typical $z\sim1-3$ galaxies.  We also aim to understand how broad emission affects the location of a galaxy on emission line diagnostic diagrams such as the N2-BPT and S2-BPT diagrams.

This paper is structured as follows:  Section 2 describes the sample, observations, data reductions, and measurements of physical properties.  Section 3 describes how we fit galaxy spectra to measure the broad emission.  We also describe how we make stacks and test the broad fitting technique.  In Section 4, we show measurements of the broad emission in individual spectra as well as stacks of galaxies.  In Section 5, we discuss the source of the broad component and consider several possible physical explanations for the broad flux such as low-luminosity AGN and shocks.   In Section 6, we discuss possible effects on physical measurements from including the flux from the broad component.  Conclusions are summarized in Section 7. 

Throughout this work we assume a $\Lambda$CDM cosmology with $\Omega_m = 0.3$, $\Omega_\lambda= 0.7$, and $H_0= 70$ km s$^{-1}$ Mpc$^{-1}$.  All magnitudes are given in the AB system (\citealt{1983ApJ...266..713O}). The wavelengths of all emission lines are in vacuum.

\section{Observations,  Reduction, and Galaxy Property Measurements}

\subsection{Observations, Reduction, and Sub-sample Selection}

In this work, we use the first two years of data from the MOSDEF survey (\citealt{2015ApJS..218...15K}) where we obtained near-infrared spectra for $\sim1500$ high-redshift galaxies using the MOSFIRE instrument (\citealt{2012SPIE.8446E..0JM}) on the W. M. Keck telescope.  These spectra were collected over the course of 48.5 nights from 2012-2016, and enable measurements of the rest-frame optical nebular emission lines for galaxies at  $1.37<z<3.80$.   The MOSDEF survey targets galaxies in the AEGIS, COSMOS, GOODS-N, GOODS-S, and UDS extragalactic legacy fields which have extensive ancillary data including Chandra, Spitzer, Herschel, \textit{HST}, VLA, and ground based optical/near-IR data.

One-dimensional spectra were extracted using custom IDL software called \texttt{BMEP}\footnote{Source code and installation instructions available at: \url{https://github.com/billfreeman44/bmep}}, as described in the Appendix. \texttt{BMEP} was tested with output from the MOSDEF team's custom 2D reduction, the MOSFIRE Data Reduction Pipeline\footnote{\url{https://www2.keck.hawaii.edu/inst/mosfire/drp.html}}, and the 2D optical spectra from the Keck Low Resolution Imaging Spectrometer (LRIS, \citealt{1995PASP..107..375O}, \citealt{2010SPIE.7735E..0RR}). For the MOSDEF data, both optimally weighted and unweighted spectra were extracted for each object, and we use the optimally weighted spectra for this analysis. The optimal extraction algorithm follows \cite{1986PASP...98..609H} but is modified to be able to extract fractions of pixels (see the Appendix). To determine the weighting profile, center, and width of each object, we fit a Gaussian to the profile of each object in each filter. The profile was determined by summing flux only at those wavelengths with high S/N in either the continuum or emission lines. Using high S/N areas of the spectra creates clean weighting profiles for the optimal extraction since wavelengths with little or no signal are excluded. 

Galaxies in the MOSDEF Survey are split into 3 redshift bins, $1.37<z<1.70$, $2.09<z<2.61$, and $2.95<z<3.80$ that were each observed using a different filter set in order to maximize efficiency of detecting multiple rest-optical emission lines of interest \citep[see][for details]{2015ApJS..218...15K}.  We combine the $1.37<z<1.70$ and $2.09<z<2.61$ galaxies into a single sample (hereafter the $z\sim2$ sample) because these galaxies have coverage of \Ha \ and \OIII.  There may be some evolution  between the galaxies at these two redshift ranges, but without combining them, broad emission is extremely difficult to detect.  The $2.95<z<3.80$ galaxies do not have coverage of \Ha \ and are stacked separately (hereafter the $z\sim3$ sample). 

The parent dataset contains 555 galaxies with measured redshifts, including 503 galaxies with \OIII \ detections and 394 \Ha \ detections.  We create a sub-sample where we remove galaxies for which it may be difficult to accurately measure the broad emission or have AGN (discussed below).  When cleaning the sample, we consider the \Ha \ and \OIII \ detections separately except when considering galaxy-wide effects which are mergers and AGN presence.

\begin{itemize}
\itemsep0em 
  \item We remove 50 \OIII \ and 35 \Ha \ detections where the galaxy was an IR, X-ray, or both IR and X-ray detected AGN (see \citealt{2014arXiv1409.6522C} and \citealt{2017ApJ...835...27A}).  AGN are a possible source of outflows (\citealt{2017arXiv170310255L}), and removing them allows us to isolate the effects of star formation on outflows.
  \item We remove 20  \OIII \ and 26 \Ha \ detections where the galaxy would be classified as an AGN based on $z\sim0$ optical line-ratio diagnostics  (above the \citealt{2003MNRAS.346.1055K} line in the N2-BPT diagram).  This is a conservative cutoff considering galaxies at $z\sim2$ are offset compared to $z\sim0$ galaxies (\citealt{2015ApJ...801...88S}).
  \item We remove 182  \OIII \ detections where  S/N $< 10$ for \OIII \ and 122  \Ha \ detections where  S/N $< 10$ for \Ha.
  \item We remove 22  \OIII \ and 41 \Ha \ detections  where the \OIII \ or \Ha \ emission line was on or near bright sky lines.
  \item We remove 14  \OIII \ and 17 \Ha \ detections where the galaxy appears to be undergoing a merger as indicated from the images or spectra of nearby objects.  Mergers may have complicated kinematics and may not be well fit by our fitting method (described in Section 3.1).  We determined which galaxies were mergers by inspecting both images and spectra by eye.  If the galaxy was very misshapen or there was a nearby companion we removed it.  In the spectra, if there were two profiles that overlapped then those galaxies were removed.
  \item We remove 7  \OIII \ and 8 \Ha \ detections  where the \OIII \ or \Ha \ emission is near the edge of wavelength coverage and the shape of the profile is difficult to determine.
  \item We remove 5  \OIII \ and 7 \Ha \ detections  for which we measured a $\text{FWHM} > 275$ from a single Gaussian.  Some galaxies have broad lines simply from rotation and velocity dispersion.  To isolate the broad emission, we restrict the narrow emission to have a $\text{FWHM} < 275$ km s$^{-1}$ as described in Section 3.1.  Including galaxies with $\text{FWHM} > 275$ km s$^{-1}$ creates false positives because the narrow component does not properly fit the narrow emission.
\end{itemize}

The final sample has 216 unique galaxies with 203 \OIII \ measurements and 138 \Ha \ measurements.     There are 125  galaxies with both an \Ha \ and \OIII \ detection.  We create stacks (discussed in Section 3.2) and the stacks at $z\sim2$ have an additional restriction to contain galaxies with wavelength coverage of  \Ha, \OIII, \SII, \NII, and \Hb, which results in 113 galaxies in the z $\sim2$ stack.  There are 60 galaxies in the $z\sim3$ stack.

\begin{figure*}
    \centering
    \includegraphics[width=0.98\textwidth]{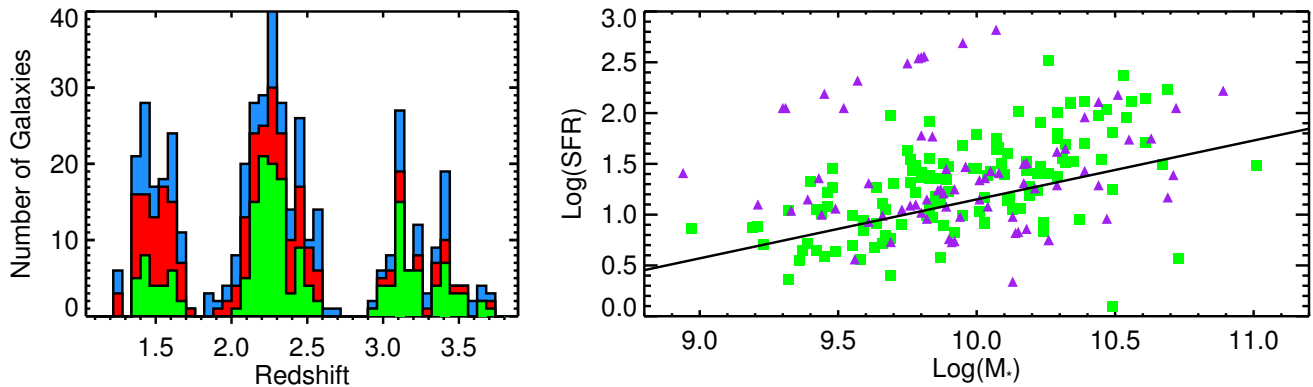}
    \caption{Left: Histograms of galaxy redshifts in the MOSDEF survey.  The solid blue histogram is the full sample, the red  is the sample after removing all galaxies with a $\text{S/N}<10$ in \Ha \  (for $z<2.3$) or in \OIII \ (for $z>2.3$), and the green histogram is the final sample.   Right: SFR vs. stellar mass for the final sample.  The green squares are $1.37<z<2.61$ galaxies and purple triangles are $2.95<z<3.80$ galaxies. The solid black line is the star-forming ``main sequence'' measured from the MOSDEF data (\citealt{2015arXiv150703017S}).}
    \label{histoplots}
\end{figure*}

Figure \ref{histoplots} shows histograms of redshift for the full sample of galaxies in blue, the remaining galaxies after the S/N rejection in red, and the the final sample in green. There is no clear bias against any particular redshift or mass, however more galaxies with $\text{SFR} <10$ \msol \ yr$^{-1}$ have been removed.  The right side of Figure \ref{histoplots} shows the SFR vs. stellar mass diagram for our sample along with the star-forming main sequence for  MOSDEF galaxies in the  $2.09<z<2.61$ range from \cite{2015arXiv150703017S}.  Because we removed galaxies that have an \Ha \ or \OIII \ $\text{S/N} <10$, our sample may be incomplete for galaxies below the main sequence and at low stellar mass.

\subsection{Stellar Population Properties}
We estimate the physical parameters for our sample, including stellar mass, SFR and age by comparing the photometric SEDs with stellar population synthesis models  (\citealt{2009ApJ...699..486C}) using the stellar population fitting code FAST (\citealt{2009ApJ...700..221K}).   We assume a \cite{2003PASP..115..763C} initial mass function (IMF) and the dust reddening curve from \cite{2000ApJ...533..682C}.  We use spectroscopic redshifts from the MOSDEF survey and broadband and mediumband photometric catalogs assembled by the 3D-HST team (\citealt{2014ApJS..214...24S})  spanning  observed  optical  to  mid-infrared wavelengths.  We include a template error function to account for the mismatch in less constrained sections of the spectrum.   For a full description of the stellar population modeling procedure see  \cite{2015ApJS..218...15K}.

When available, we derive SFRs based on the \Ha \ emission line by correcting for Balmer absorption (using the SED) and dust extinction (using the Balmer decrement of \Ha/\Hb), then converting the \Ha \ luminosity into a SFR \citep{1998ARA&A..36..189K}, adjusted for a \cite{2003PASP..115..763C} IMF (see \citealt{2015arXiv150703017S} for more details). Because galaxies in the $z\sim3.3$ bin do not have coverage of \Ha, we use SED fitting to determine their SFRs.

\section{Measuring the Broad Component}
In this section, we describe the technique for measuring the broad emission line components for individual galaxies. We also describe how we create stacks.

\subsection{Fitting galaxies}

We aim to measure an underlying broad component of emission lines of galaxies.  By assuming that emission lines are composed of narrow and broad components with Gaussian profiles. This section describes the fitting process as well as constraints on parameters.

The \Ha, \NII, and  \SII \ lines are in the same filter (H if at $z\sim1.5$ and K if at $z\sim2.3$) while  \OIII \ and \Hb \ fall into a different filter (J if at $z\sim1.5$, H if at $z\sim2.3$, and K if at $z\sim3.3$).  We do not simultaneously fit lines that are in different filters because the spectral resolution is slightly different in each filter\footnote{\url{www2.keck.hawaii.edu/inst/mosfire/grating.html}}.  Additionally the seeing may vary between different filters. We do not include \SII \ in fits of individual galaxies because it is too faint to measure the broad component, however it is included in fits of stacks (see Section 3.2). For individual galaxies at $1.37<z<2.61$ we fit \Ha \ and \NII~$\lambda\lambda$6549,6583 simultaneously.  For individual galaxies at $1.37<z<3.8$ we fit  [\ion{O}{3}]~$\lambda$5008, [\ion{O}{3}]~$\lambda$4959,  and \Hb \ simultaneously.

For each set of lines, we perform two preliminary fits and one final fit.  The first preliminary fit uses a single Gaussian to fit each emission line using \texttt{MPFIT}, a non-linear least squares fitting code \citep{2009ASPC..411..251M}.  We use this single Gaussian fit to subtract off a linear continuum and normalize the data so the fitted flux density of the peak of the brightest line for each set of lines (\Ha \ or \OIII) is unity, respectively.  Next, we fit the data again using \texttt{MPFIT} but this time each emission line is fit with two Gaussians, one broad and one narrow.  We use the resulting values and errors of this second fit as initial values for the final fit which is done with a custom Markov Chain Monte Carlo (MCMC) code \texttt{MPMCMCFUN}\footnote{Source code and installation instructions available at: \url{https://github.com/billfreeman44/mpmcmcfun}} that uses the Metropolis-Hastings algorithm (\citealt{doi:10.1063/1.1699114}, \citealt{doi:}).  The errors in the second preliminary fit are used as the parameter jump amplitudes for the final fit.  The final MCMC fit is necessary because it offers a better characterization of errors.  This is especially relevant in these fits because we are fitting one emission line with two Gaussian components and the correlation between parameters may be significant. In what follows, we use the subscripts S, B, and N  to distinguish parameters for the single, broad, and narrow components, respectively.

When fitting multiple emission line components, we constrain the FWHM$_{\text{B}}$, FWHM$_{\text{N}}$, broad component shift ($\Delta$v), constant background, and the narrow component redshift to be the same for each line. This leaves each single emission line with two free parameters, broad amplitude (A$_{\text{B}}$) and narrow amplitude (A$_{\text{N}}$).  Two exceptions to this are the  \NIII / \NIIII \ and [\ion{O}{3}]~$\lambda$5008/ [\ion{O}{3}]~$\lambda$4959 flux ratios which are set to $2.93$ and $2.98$ respectively according to atomic physics (\citealt{1989agna.book.....O}).  Therefore, each fit has five free parameters shared by each line (FWHM$_{\text{B}}$, FWHM$_{\text{N}}$, $\Delta$v, narrow component redshift, and constant background) and two free parameters for each line (A$_{\text{N}}$ and A$_{\text{B}}$).

The resulting best-fit parameters are likely to depend on the chosen limits. For instance, not placing a minimum on the FWHM$_{\text{N}}$ can result in an unphysically narrow emission line. Also, not placing a minimum on the FWHM$_{\text{B}}$ can result in the broad component not being representitive of broadened emission. Therefore, we place physically motivated restrictions on all free parameters.  For individual galaxies, we restrict the FWHM$_{\text{N}}$ so that it cannot be lower than the average FWHM of skylines in that particular filter and mask. For galaxies smaller than the slit width, it is possible that the FWHM of the narrow component is smaller than that of skylines. In most cases, the seeing is not smaller than the width of the slit (0\farcs7).  Emission lines are also broadened by the spatial extent of the galaxy and the velocity distribution therein. Therefore, we do not expect FWHM$_{\text{N}}$ to be much lower than the width of the sky lines.  We also restrict the FWHM$_{\text{N}}$ to be less than 275 km s$^{-1}$. For this sample, we have removed galaxies where FWHM$_{\text{S}}$ is larger than 275 km s$^{-1}$ (see Section 2.1).

In order to properly study outflows, we must be certain that the broad components measure a kinematically distinct feature from the rotation of the host galaxy.  In other words,  the broad flux must not be an artifact from a better fit to the narrow emission by using two Gaussian components. Therefore, we restrict the minimum FWHM$_{\text{B}}$ to be a larger value than could be reasonably fit using only a single Gaussian component. Accordingly, we set the minimum FWHM$_{\text{B}}$ to be 300 km s$^{-1}$ which provides some separation in the velocities of the narrow and broad components.  Typical FWHMs for ionized outflows from star-forming galaxies are 300-600 km s$^{-1}$ (\citealt{2012ApJ...761...43N}, \citealt{2011ApJ...733..101G}, \citealt{2015arXiv150700346W}). 
Some studies have measured galactic outflow speeds $>1000 \ \text{km s}^{-1}$, but these are typically associated with AGN (\citealt{2009ApJ...701..955S}, \citealt{2014ApJ...796....7G}, \citealt{2014ApJ...787...38F}).  Since we have removed known AGN, we do not expect outflows of such high velocity.  The upper limit of FWHM$_{\text{B}}$ is set to 850 km s$^{-1}$ which is the typical maximum velocity deduced by the blue-shifted interstellar absorption lines in the rest frame UV of $z\sim2$ galaxies (\citealt{2010ApJ...717..289S}).

When the amplitude of the broad component is consistent with zero, the center and width become unconstrained.  Therefore it is important to restrict these parameters so they do not stray to unrealistic values.  The centroid of the broad component is allowed to be anywhere within $\pm 100 $ km s$^{-1}$ of the expected value.  The broad component shift is the same for each line.  No objects that had significant detections of the broad component ran into this limit.  Other studies typically find shifts of $<100$ km/s (\citealt{2012ApJ...761...43N}, \citealt{2015arXiv150700346W}).

The amplitudes of the narrow  \Ha \ and \OIII \ components are constrained to be between 0.2 and 1.05 (the peak of these lines were normalized to unity from the single Gaussian fit) and the broad component amplitude is constrained to -0.3 and 0.8.  Since some galaxies do not show any signs of broad emission, the best value for the FWHM$_{\text{B}}$ could be at or near zero for these galaxies.  In these cases, it is still useful to put limits on the broad emission.  Therefore, we allow the FWHM$_{\text{B}}$ to be negative to fully sample the parameter space and set proper upper limits on the FWHM$_{\text{B}}$.  In cases where the best value for the FWHM$_{\text{B}}$ is less than zero, we interpret this galaxy as having no significant broad emission but still show the upper limit.  For \NII, \SII, and \Hb \ we scale the restrictions to the relative peak of each line.  All of the constraints are listed in Table \ref{constraints}.

\begin{deluxetable}{lll} 
\tablecaption{Constraints of the Fits}
\tablewidth{0pt} 
\tablehead{
\colhead{Parameter} &
\colhead{Minimum} &
\colhead{Maximum} 
}
  \startdata
FWHM$_{S}$\tablenotemark{a} & varies\tablenotemark{b}  & 275  \\
A$_{S}$\tablenotemark{c} & 0 & 1.05\\
$\Delta \text{v}_{S}$\tablenotemark{d} & -100  & +100  \\ 
FWHM$_{\text{N}}$ & varies\tablenotemark{b}  & 275 \\
A$_{\text{N}}$ & 0.2  & 1.05 \\
$\Delta \text{v}_{\text{N}}$ & -100  & +100  \\
FWHM$_{\text{B}}$ & $300$ & 850 \\
A$_{\text{B}}$ & -0.3 & 0.8 \\
$\Delta \text{v}_{\text{B}}$ & -100  & +100
 \enddata
\tablenotetext{a}{FWHM in units of km s$^{-1}$}
\tablenotetext{b}{Set to the average FWHM of skylines in each mask}
\tablenotetext{c}{Relative to the maximum flux of the line}
\tablenotetext{d}{Center shift in units of km s$^{-1}$. Negative values imply blueshifts}
\label{constraints}
\end{deluxetable}

\begin{figure*}
    \centering
    \includegraphics[width=0.98\textwidth,height=0.98\textheight]{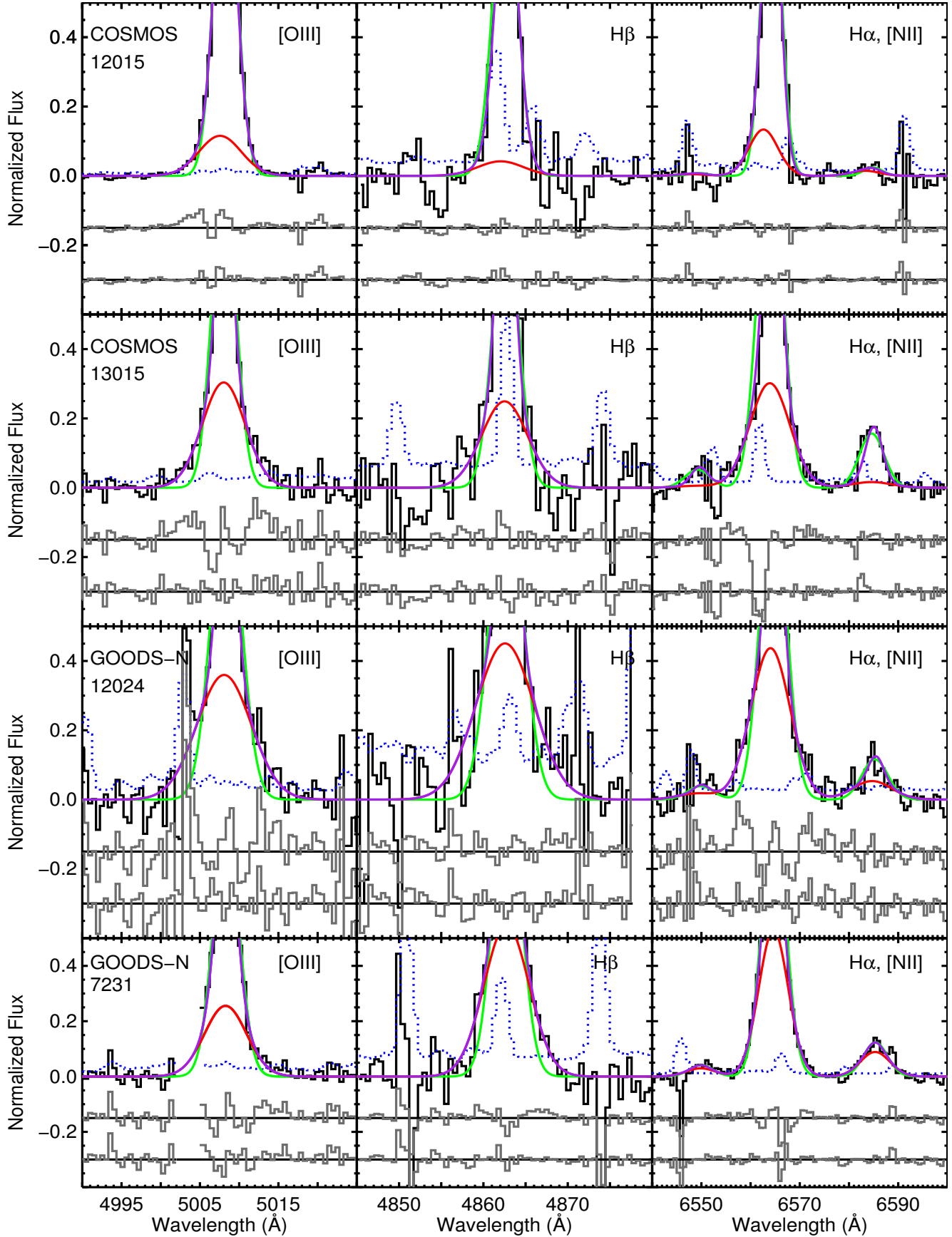}
    \caption{Four example fits for individual spectra.  Each row is one object and each column from left to right is \OIII, \Hb, and \Ha. The field and 3D HST v2.1 catalog ID is in the upper left.  Each line is normalized such that the strongest line peak is unity.  The single Gaussian fit is shown in green.  The overall fit for the narrow+broad fits for each stack is shown in purple with the broad component for this fit shown in red.  The error spectrum is shown as a dotted blue line.  The two gray lines show the residuals for the single Gaussian fit (top) and the narrow+broad fit (bottom).  The horizontal solid black lines show the amount each residual is offset.  A skyline has been masked for GOODS-N 12024 at 5004 \AA \ and at 4877 \AA.}
    \label{aaaplot}
\end{figure*}

Figure \ref{aaaplot} shows fits for four galaxies that exhibit the strongest evidence for a broad component.  It is clear that a single Gaussian does not fit these galaxies well as evidenced by  the ``wave" pattern that is present in the residuals.  The pattern shows the single fits underestimate flux at the peak, overestimate the wings, and underestimate the base. This pattern is particularly evident in the \OIII \ lines of COSMOS-12015, and COSMOS-13015 and in the \Ha \ lines of GOODS-N-12024 and GOODS-N-7231.  

The broad flux only dominates a small fraction of the line at high velocities.  If the broad flux component is not approximately Gaussian, then the fitted broad flux might be different from what we measure.  Other studies that analyze galaxies with higher signal to noise find that the broad emission is typically well fit by a Gaussian (\citealt{2012ApJ...761...43N}, \citealt{2009ApJ...701..955S}, \citealt{2011ApJ...733..101G}).

\subsection{Making Stacks of Spectra}

\begin{figure*}
    \centering
    \includegraphics[width=0.98\textwidth]{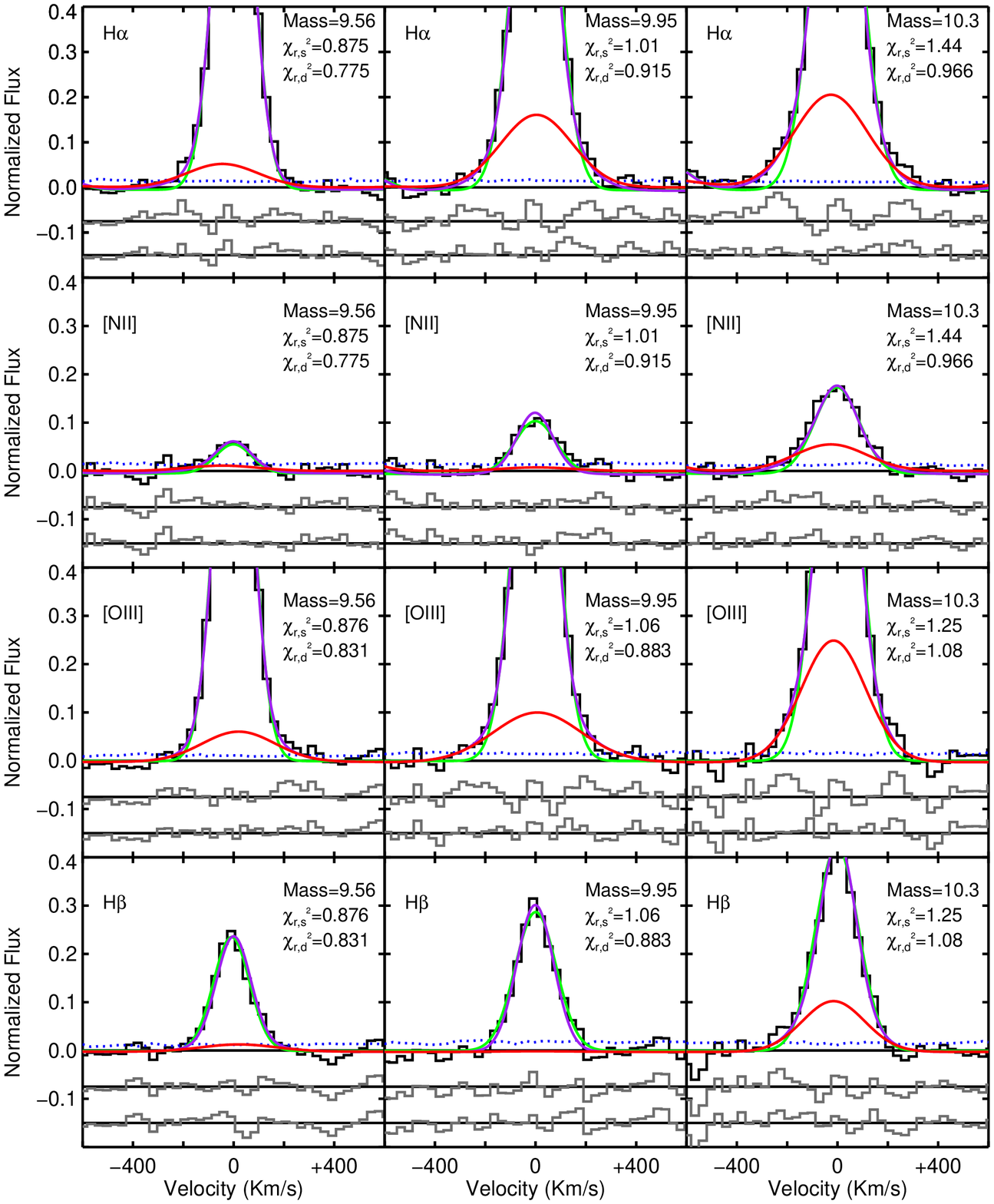}
    \caption{Stacks of galaxies showing both the single Gaussian and narrow+broad component fits. The rows show \Ha ,\NII, \OIII, and \Hb \ lines from top to bottom.  The columns show each stack with the stellar mass increasing from left to right. The line colors have the same meaning as in Figure \ref{aaaplot}. $\chi_{r,s}^2$ is the reduced $\chi^2$ for the single fit and The $\chi_{r,d}^2$ the reduced $\chi^2$ for the narrow+broad fit.}
    \label{stacks-gals-combined}
\end{figure*}

The broad component is difficult to separate from the narrow emission. The galaxies in Figure \ref{aaaplot} were chosen because they show the strongest evidence for broad emission.  The faint, high velocity wings of the broad component are difficult to distinguish from the noise for the majority of the individual galaxies.   In order to achieve a higher S/N, we create stacks of galaxies in bins of stellar mass such that each stack has approximately the same number of galaxies. To create each stack, we interpolate the flux for each galaxy to a common rest-frame wavelength grid, subtract off any continuum, convert each spectrum from flux density to luminosity density, divide by the total  luminosity of either \Ha \ or \OIII \ depending on the wavelength coverage of the stack, and then sum each spectrum with no weighting. To avoid adding significant noise from sky line subtraction residuals, we remove pixels associated with sky lines where the error spectrum is above 1.5$\times$ the median error. The error in the stacked spectrum is calculated by making the stack 200 times but using input spectra with added Gaussian noise according to the associated error spectrum for each individual object; the error is calculated by taking the standard deviation of the 200 stacks at each wavelength. The $z\sim2$ stacks are shown in Figure \ref{stacks-gals-combined}.

With this sample, it is also possible to create stacks in bins of SFR, sSFR, and $\Sigma_{SFR}$.  These stacks are not truly independent from the stacks by mass because these properties correlate with mass. We chose to use mass for the primary analysis in this work because, of the physical parameters we considered, mass is the only parameter that is not estimated by using the \Ha \ emission line which may be influenced by broad emission.

The line fitting process for stacks is the same as described in Section 3.1 with some exceptions. The doublet \SIIl \ is included when fitting \Ha \ and \NII. For the lower limit of FWHM$_{\text{N}}$, we use the average skyline FWHM for each galaxy, which is $80$ km s$^{-1}$. The fit to each stack is plotted in Figure \ref{stacks-gals-combined}. This figure shows \Ha \ and \NII \ for the stacks at $1.37 < z < 2.61$.  For both \Ha\ and \OIII \ in all stacks, the amplitude of the broad component is significant at the $>3\sigma$ confidence level.

\subsection{Assessment of False Positives}
The broad component dominates the line only at the highest velocities, which is also where the S/N is the lowest. Here we test the fitting process to show that measured broad line parameters are consistent with simulated input parameters.

For this test, we take a single Gaussian, add noise, and fit the Gaussian using the method described in Section 3.1.  Since this idealized Gaussian has no actual broad component, any broad component that we measure is a false positive. We performed this test on 200 simulated emission lines, each with 10 different FWHMs between 75 and 275 km s$^{-1}$, which span the range of measured narrow components from the MOSDEF sample.  We used the same resolution and wavelength as for an \Ha \ line for a $z\sim2.3$ galaxy.  We add a constant amount of noise to each spectrum such that the S/N of the \Ha \ line ranges between 10 and 300. The assumption of constant noise is an appropriate approximation of the error for a single emission line in the actual spectra because we are limited by the bright sky rather than Poisson noise from the object. The noise does increase as a function of wavelength, particularly in the K band, but this test was only on a single emission line and the error does not change much over the span of a single emission line.

We find that only  11\%, 1\%, and 0.05\%  of simulated galaxies have a false positive of 1$\sigma$, 2$\sigma$, or 3$\sigma$ respectively.  These are lower than the expected rates based on Gaussian statistics, which are 16\%, 2\%, and 0.1\%.  The slightly lower values of the test result are because we allow the narrow peak to exceed 1.0 (the max is 1.05).  The average broad component is slightly less than 0 which creates an offset in the number of 1$\sigma$, 2$\sigma$, or 3$\sigma$  detections.  Another result of this test is that the fraction of false positives did not change as a function of width or noise added.

In addition to false positives in individual spectra, there is a possibility of creating an artificial broad component when making the stacks.  We performed several tests to ensure that in creating the stacks, we did not also create an artificial broad component.  The first test takes 50 Gaussians of random FWHM between 75 and 250 km s$^{-1}$, adds noise, and creates a stack as described in Section 3.2.  Each added Gaussian has a constant amount of noise across each wavelength element which is similar to the level in actual spectra except for skylines which are not included in these simulated stacks.  We found no evidence of introducing false positives when creating stacks.  We repeat this test and add a random shift between $\pm 1$ \AA \ to the centroid of the Gaussian which simulates imperfect redshift estimates.  The average redshift error for this sample is $6\times10^{-5}$ which is $\sim 0.13$ \AA \ at these redshifts. These stacks also failed to produce false positives. 

These tests have shown that we do not expect false positives to be an issue when using the fitting method described in Section 3.1.  We have also shown that creating stacks of galaxies does not introduce a broad emission signature.

\section{Results} 

In this section, we discuss the broad flux measured in individual and stacked spectra.  We discuss the physical interpretation of these measurements in the subsequent section.

\subsection{Broad Flux Ratio}
After fitting each galaxy and stack, we parameterize the broad emission we measured as a broad to narrow flux ratio (broad flux / narrow flux, BFR).  We chose this parameterization because other studies have used this and using the same parameterization allows for easy comparision (e.g. \citealt{2012ApJ...761...43N}).  The BFR is also used to estimate the mass loading factor (Section 5.3). The other natural parameterization, broad flux to total flux, can be  calculated as $\text{(broad flux/total flux)} = 1/(1 + \text{BFR}^{-1})$.

\begin{figure*}
    \centering
    \includegraphics[width=0.98\textwidth]{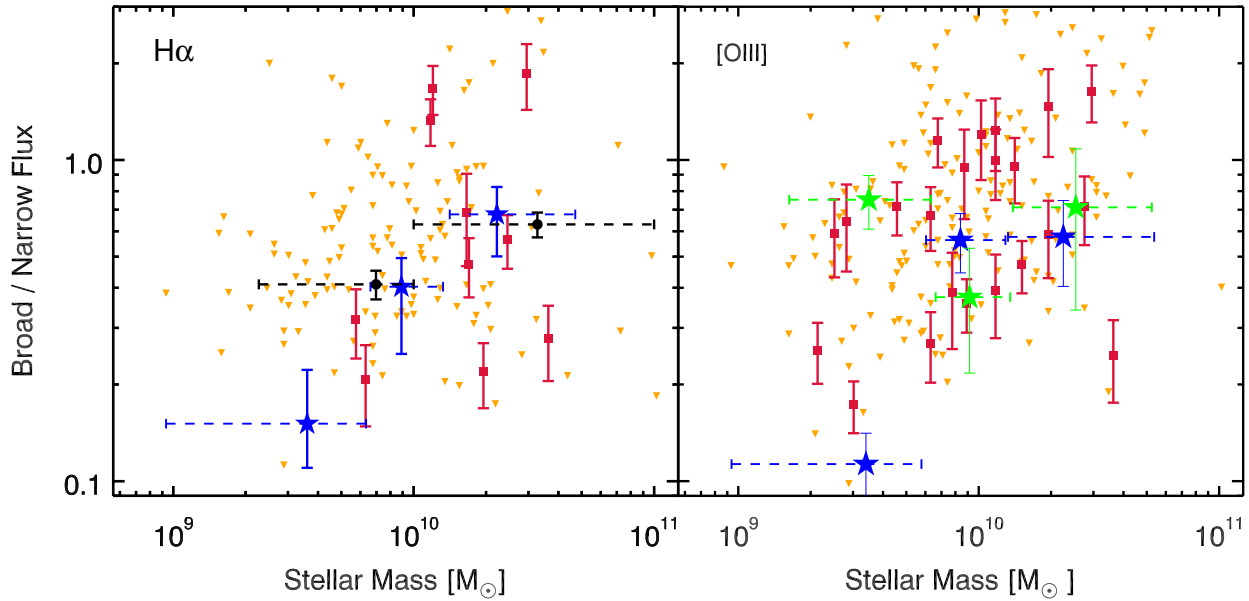}
    \caption{The BFR as a function of mass for \Ha \ (left) and \OIII \ (right).  Red squares are galaxies with a broad component detection of $>3\sigma$ significance with $1\sigma$ error bars plotted.  Orange triangles are $3\sigma$ upper limits for galaxies with $<3\sigma$ significance.   Blue stars show the BFR of the $z\sim2$ stacks, and the green stars are the $z\sim3$ stacks.  The black circles are stacks from \cite{2012ApJ...761...43N}.  For the stacks, the vertical error bars are $1\sigma$ error bars from the fit and the horizontal dashed lines show the range of points included.}
    \label{quadplot}
\end{figure*}

\begin{figure*}
    \centering
    \includegraphics[width=0.98\textwidth]{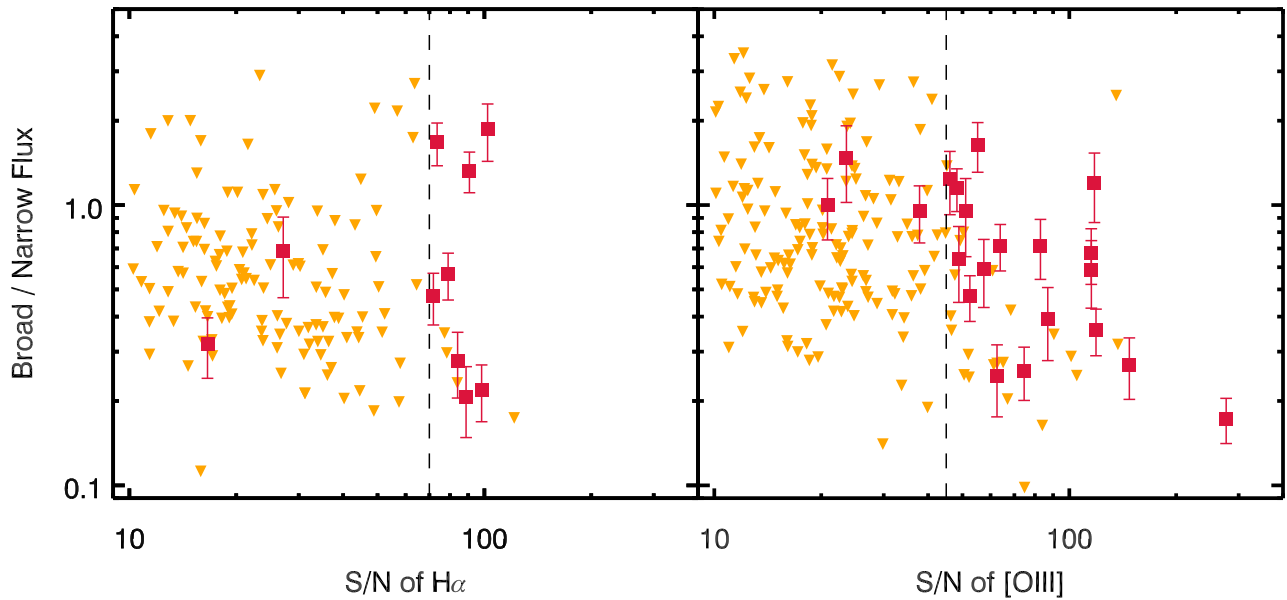}
    \caption{The BFR as a function of S/N for \Ha \ (left) and \OIII \ (right).  Red squares are galaxies with a broad component detection of $>3\sigma$ significance with one sigma error bars plotted.  Orange triangles are $3\sigma$ upper limits for galaxies with $<3\sigma$ significance. Vertical lines are drawn at S/N=70 for \Ha \ and S/N=45 for \OIII.  For \Ha, 66\% of galaxies with S/N $>70$ have broad component detections but only 1.6\% of galaxies with S/N $<70$ have detections.  For \OIII, 32\% of galaxies with S/N $>45$ have broad component detections but only 5\% of galaxies with S/N $<45$ have detections.  The location of the vertical lines was chosen by eye to emphasize the dependence of detecting broad flux and S/N.  The dependence of the detection of the broad flux on S/N implies that the 10\% detection rate is a lower limit.}
    \label{sn}
\end{figure*}

The left side of Figure \ref{quadplot} shows the BFR measured from the  \Ha \ line as a function of mass.  For individual galaxies, there are 10 detections with $>3\sigma$ significance out of 138 galaxies (7\%).  For the stacks and  the galaxies with detections, the broad flux accounts for 10-70\% of the total flux in nebular emission lines.  The MOSDEF measurements for BFR are consistent with the measurements from \cite{2012ApJ...761...43N} who did a similar analysis for galaxies at the same mass range.  The details of the fits are in Table \ref{tablehalpha}. The small differences in the number of significant detections in this study and \cite{2017arXiv170310255L} for the same sample can be attributed to slight differences in codes used to fit the data.

The right side of Figure \ref{quadplot} shows the BFR measured from the  \OIII \ lines as a function of mass. For \OIII \ there are 21 detections with $>3\sigma$ significance out of 201 galaxies (10\%).  For the stacks and galaxies with detections, the broad flux accounts for 20-50\% of the total flux in nebular emission lines.  For the $z\sim3$ stacks the broad component comprises 30-60\% of the flux and the BFR is slightly higher than in the  $z\sim2$ stacks on average.

The stacks in Figure \ref{quadplot} show an apparent correlation between the BFR and mass.  Measuring the broad flux in the lowest mass stack is difficult because most of the broad emission may be at $\text{FWHM} < 275 \ \text{km s}^{-1}$ and reliably detecting low velocity broad emission is difficult (discussed in detail in Section 5.3).  Non-measurement of the broad fluxes for the lowest mass galaxies may introduce a bias in the BFR vs. stellar mass relation.  Additionally, the \OIII \ broad emission at $z\sim2$ does not show any increase above $7\times10^9$ \msol, and the $z\sim3$ stacks show no change with mass.  There is also no correlation between the detected broad emission in individual glaxies and mass.  For these reasons, we cannot confirm a correlation between the BFR and mass.

Figure \ref{sn} shows the BFR measured from the  \Ha \ line as a function of S/N.  Here, we define the S/N as the fitted flux by a single Gaussian divided by the error in the flux for either \Ha \ or [\ion{O}{3}]~$\lambda$5007.  Galaxies with detections tend to also be at higher S/N.  For \Ha, 66\% of galaxies with S/N $>70$ have broad component detections but only 1.6\% of galaxies with S/N $<70$ have detections.  For \OIII, 32\% of galaxies with S/N $>45$ have broad component detections but only 5\% of galaxies with S/N $<45$ have detections.  These two thresholds were chosen by eye to emphasize the S/N dependence on detecting the broad component.  It is easier to detect broad emission in \OIII \ than in \Ha \ because we include the  [\ion{O}{3}]~$\lambda$4959 emission when fitting [\ion{O}{3}]~$\lambda$5007.  Using both lines in the fit provides a better constraint on the shape of the broad and narrow emission profiles than using only one line.

The dependence of the detection of the broad flux on S/N implies that the 10\% detection rate is a lower limit.  Because outflows are supposedly ubiquitous at $z\sim2$ we would likely see more broad component detections with deeper data.  A dependence on S/N and a broad component detection was also seen in \cite{2017arXiv170310255L} for AGN in the MOSDEF sample.

\subsection{Broad and Narrow Component Line Ratios}

\begin{figure*}
    \centering
    \includegraphics[width=0.98\textwidth]{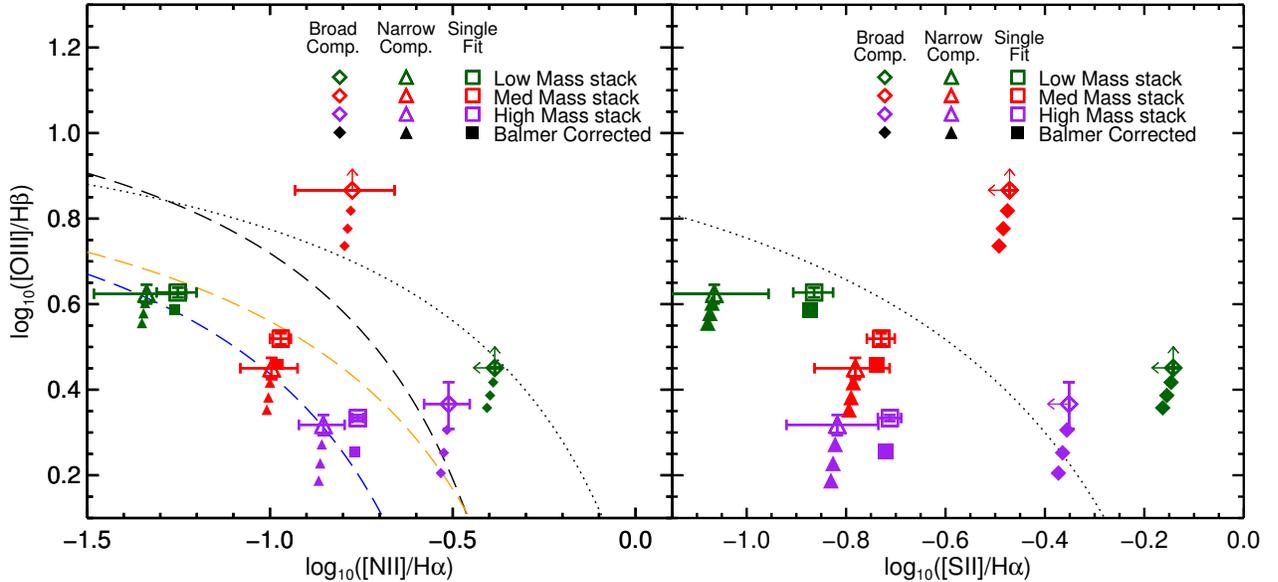}
    \caption{The N2-BPT and S2-BPT diagram for $z \sim2.3$ stacks of data by mass.  The ratios for each stack were calculated using the narrow component, broad component, and single Gaussian fits.  The broad component ratios, narrow component ratios, and single Gaussian fit ratios are the diamonds, triangles, and squares respectively.  The solid points are corrected for Balmer absorption.  For the narrow and broad line ratios, the three points show if 33, 66, and 100\% of the Balmer absorption is applied to that particular component.  Error bars are 1$\sigma$ and galaxies with $\text{S/N}<3$ are plotted at 1$\sigma$ limits and are marked by arrows. The blue dashed line is measured from \cite{2013ApJ...774..100K} for local galaxies.  The orange dashed line is measured from \cite{2015ApJ...801...88S} for $z\sim2.3$ galaxies in the MOSDEF survey.  The dashed black line is the line from \cite{2003MNRAS.346.1055K} that separates star forming galaxies and AGN.  The dotted black line is from \cite{2001ApJ...556..121K} and is the ``maximum starburst" line where above this line lie AGN.}
    \label{bptdiagram}
\end{figure*}

As described in Section 3.1, we fit narrow and broad components to the \OIII, \Hb, \Ha, \NII, and \SII \ emission lines in stacked spectra.  From this analysis, we are able to calculate the \NII/\Ha, \SII/\Ha, and  \OIII/\Hb \ ratios and place each component on the N2-BPT and S2-BPT diagrams.  Figure \ref{bptdiagram}  shows the N2-BPT and S2-BPT diagram for the low, medium, and high mass stacks.  We do not include individual galaxies here because there were not enough $3\sigma$ detections of the broad components of \Hb, \NII, and \SII \ and we could not create robust line ratios.  The blue dashed line is measured from \cite{2013ApJ...774..100K} for local galaxies.  The orange dashed line is measured from \cite{2015ApJ...801...88S} for $z\sim2.3$ galaxies in the MOSDEF survey.  The dashed black line is from \cite{2003MNRAS.346.1055K}, and separates star forming galaxies and AGN in the local universe.  The dotted black line is from \cite{2001ApJ...556..121K} and is the ``maximum starburst" line where galaxies containing AGN lie above this line.

For individual galaxies, Balmer emission line fluxes can be corrected for underlying stellar absorption based on the equivalent widths of stellar Balmer features as estimated from the stellar population synthesis model fit to the SED of each galaxy (\citealt{2015arXiv150402782R}).  For each stack, we estimate the Balmer absorption by calculating the average absorption for each galaxy in the stack.  This gives us an estimate for the total fraction of flux that was absorbed but no information about the shape.  Without knowing the exact shape/width of the absorption feature, we do not know how much of the correction should be applied to the narrow feature and how much should be applied to the broad feature.  Therefore, we calculate the Balmer absorption correction assuming the broad component is affected by 0, 33, 66, and 100\% of the Balmer absorption and the narrow component is affected by 100, 66, 33, and 0\% respectively.  This gives a general idea of the most the Balmer absorption could affect each line ratio.  In Figure \ref{bptdiagram}, the 0\% and 100\% absorption cases correspond to the hollow point and the solid point furthest from the hollow point, respectively.   For the single Gaussian fits (square points) there is only one solid point because the Balmer emission is not split between narrow and broad components and the magnitude of the Balmer absorption correction is unambiguous.  The shape of the Balmer absorption may also affect the fits in a manner that is difficult to predict.

In Figure \ref{bptdiagram}, the ratios from the single Gaussian fits (squares) are lower than results from previous MOSDEF studies (\citealt{2015ApJ...801...88S}).  This can be explained by the fact that we required a $\text{S/N}>10$ for the \Ha \ and \OIII \ lines.  This requirement preferentially removed lower mass galaxies which are typically more offset from the local relation (\citealt{2015ApJ...801...88S}).

The narrow component ratios tend to lie more towards the $z\sim0$ relationship (blue dashed line, \citealt{2013ApJ...774..100K}) than other measurements at $z\sim2.3$.  The broad components of the narrow+broad fits (diamonds) lie in the composite region or above the \cite{2001ApJ...556..121K} line.  The Balmer correction is large for the  \OIII/\Hb \ ratio  and this makes it difficult to conclude if the broad components have higher \OIII/\Hb \  than their narrow counterparts.  The Balmer correction is smaller for the \NII/\Ha \ ratio and it  is clear that the broad components have higher \NII/\Ha \ than the narrow components even after Balmer absorption correction.

The right side of Figure \ref{bptdiagram} shows the S2-BPT diagram for each stack and for each component.  The  \SII \ line typically has less flux than the \NII \ line making measuring the broad component more difficult.  We are only able to place $1\sigma$ limits on the broad components of the \SII \ line in stacks.   Nevertheless, these limits are consistent with a higher  \SII/\Ha \ ratio for the broad components.

In addition to the fits, we calculated line ratios by integrating the flux in several velocity bins with respect to the centroid of the line.     This measurement provides a non-parametric estimation of the line ratios that is independent of any model and is shown in the Appendix. We found that the higher velocity bins generally had higher \NII/\Ha \ and \OIII/\Hb \ ratios which is consistent with what we measure with the fits.

\section{Discussion}

In this section we discuss possible origins of the broad flux emission that can explain the offset line ratios of the broad component compared to the narrow component.  We consider shocks (Section 5.1) and low luminosity AGN (Section 5.2).  We also interpret the broad emission as an outflow and estimate the mass loading factor for the stacks (Section 5.3).

\subsection{Shocks}

Emission line ratios from shocks differ from ratios in photoionized gas.  Shock-heated gas can become ionized by high-energy photons from the shock or excited by collisions.   Emission line ratios shift in the presence of shocks and the magnitude and direction of the shift depends on the metallicity, electron density, magnetic field, and shock velocity (\citealt{2008ApJS..178...20A}).  Shocked emission tends to have higher \NII/\Ha \ and \SII/\Ha \ ratios relative to what is produced in photoionized HII regions (\citealt{2008ApJS..178...20A}).  Since the broad components in Figure \ref{bptdiagram} have higher \NII/\Ha \  ratios than the narrow components or single Gaussian fits, this may indicate the presence of shocks.  In this section, we investigate if the broad emission can be explained by shocks by creating the N2-BPT and S2-BPT diagrams using data from the shock models by \cite{2008ApJS..178...20A}\footnote{\url{http://cdsweb.u-strasbg.fr/~allen/shock.html}} and comparing these models to the broad emission line ratios.  

The shock models simulated emission line ratios for shocked gas, the precursor to the shock, and a shock+precursor which combines the shock and precursor components.  The precursor is material that is photoionized by the shock but not directly shocked itsself.  Because we do not spatially resolve the emission from these galaxies, we are unable to separate the different components of the shock.  Therefore, we compare our measurements of the broad emission to the combined shock+precursor ratios.

\cite{2008ApJS..178...20A} measured shock+precursor emission line ratios for two sets of models, one at a fixed electron density with varying metallicity ($\text{n}_{\text{e}}=1 \ \text{cm}^{-3}$ at log(O/H)+12 of 8.03, 8.35, 8.44, and 8.93), and another at fixed metallicity with varying electron density ($\text{log(O/H)}+12=8.93$ at $\text{n}_{\text{e}}=1, 10, 100, 1000 \ \text{cm}^{-3}$).  We restrict the models shown to those that have a  magnetic field strength at pressure equipartition. The shock velocity for the models range from $100-1000 \ \text{km s}^{-1}$, but we only show shock velocities of 200-500 $\text{km s}^{-1}$ based on the velocities measured in Table 2.  

\begin{figure*}
    \centering
    \includegraphics[width=0.98\textwidth]{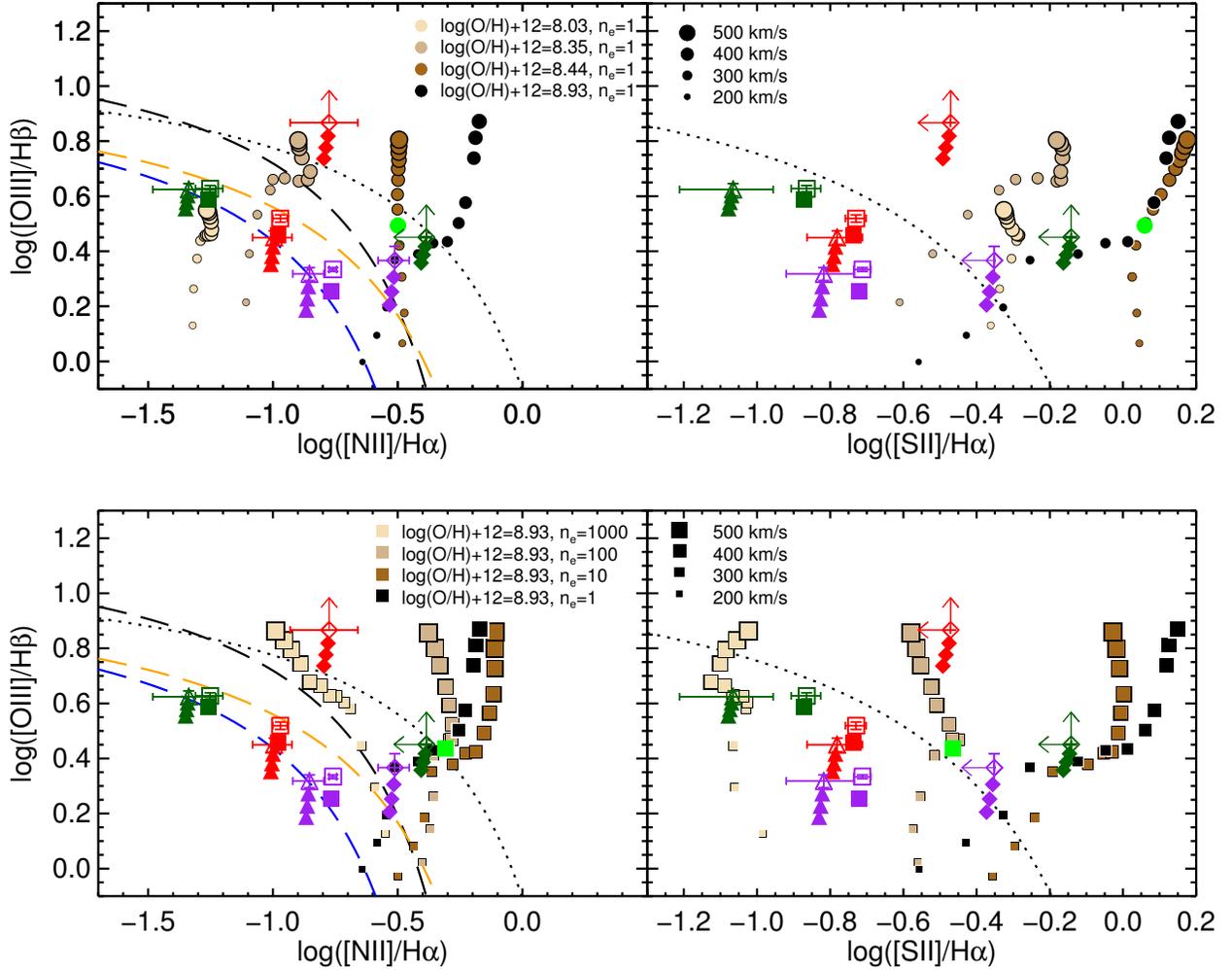}
    \caption{Shock models by \cite{2008ApJS..178...20A} for the N2-BPT and S2-BPT diagrams overlaid on the line ratios  measured from the stacks which are shown using the same symbols as in Figure \ref{bptdiagram}.  The models in the top row change in metallicity and the models in the bottom row change in electron density (in the legend, the units for electron density are $\text{cm}^{-3}$).  The green point corresponds to the shock model which has ($v=300$ km s$^{-1}$, log(O/H)+12 = 8.44, $\text{n}_{\text{e}}=1 \ \text{cm}^{-3}$) in the top row and ($v=300$ km s$^{-1}$, log(O/H)+12 = 8.93, $\text{n}_{\text{e}}=100 \ \text{cm}^{-3}$) in the bottom row.  This point is the best match to the metallicity, electron density, and shock velocity for the entire sample.}
    \label{shocksbpt}
\end{figure*}

In Figure \ref{shocksbpt}, we show the shocked models for the N2-BPT and S2-BPT diagrams.  The top row shows the effect of changing metallicity on shocked diagnostic ratios, and the bottom row shows the effect of changing density.  The galaxies in this sample (with S/N of \NII$>3$) have a median metallicity of $\text{log(O/H)}+12=8.43$ with 80\% of galaxies between $8.27<\text{log(O/H)}+12<8.59$ calculated using the \NII/\Ha \  ratio as in \cite{2015ApJ...799..138S}. The electron density of the MOSDEF galaxy sample at $2.09<z<2.61$ is 290 $^{+88}_{-169}$ cm$^{-3}$ (\citealt{2016ApJ...816...23S}).   This was calculated using the entire \SII \ line and the electron density of the material causing the broad emission may be different.  Newman et al. (2012) measured a density of  $10$ $^{+590}_{-50}$ cm$^{-3}$ from a stack of 14 galaxies, and this value is consistent with our assumption of 290 cm$^{-3}$.  Since none of the simulations span exactly the range of metallicities and densities of the MOSDEF galaxies, we are forced to extrapolate between the effects of metallicity and density.  We highlight the point that is the best match to the metallicity, electron density, and shock velocity in green.  This green point corresponds to the shock model which has ($v=300$ km s$^{-1}$, log(O/H)+12 = 8.44, $\text{n}_{\text{e}}=1 \ \text{cm}^{-3}$) in the top row and ($v=300$ km s$^{-1}$, log(O/H)+12 = 8.93, $\text{n}_{\text{e}}=100 \ \text{cm}^{-3}$) in the bottom row.

In Figure \ref{shocksbpt}, there is a strong metallicity dependence on the \NII/\Ha \ ratio and there is almost no change as a function of electron density except at the highest density ($\text{n}_{\text{e}}=1000 \ $cm$^{-3}$) where \NII/\Ha \ decreases.  The broad lines measured from stacks are consistent with the [log(O/H)+12=8.44, $\text{n}_{\text{e}}=1$ cm$^{-3}$] and [log(O/H)+12=8.35, $\text{n}_{\text{e}}=1$ cm$^{-3}$] points.  The shock model that best matches the physical parameters (green point) is very near the broad emission line ratios.  Therefore, it is feasible that the positions of the broad components in the N2-BPT diagram can be explained by shocks.

In the top row of Figure \ref{shocksbpt}, the model that best matches the MOSDEF data has a higher  \SII/\Ha \ ratio than any of the broad emission.  This may be due to the limitation that the models with varying metallicity have an electron density of 1 cm$^{-3}$.  The models in the bottom row show that \SII/\Ha \ decreases as electron density increases.  Since the electron density of the MOSDEF galaxy sample is 290 $^{+88}_{-169}$ cm$^{-3}$ the models in the top row would likely shift to lower \SII/\Ha \ ratios at higher electron densities.  Therefore, it is possible that the positions of the broad components in the S2-BPT diagram can be explained by shocks.

From these line ratios, we conclude that it is possible that the broad emission is a result of shocked emission.  This explaination does not rule out other sources of emission such as AGN (discussed in Section 5.2) and photoionized outflows (discussed in Section 5.3).

\subsection{AGN}

When creating the sample presented in this work, we removed all X-ray, IR, and optically identified AGN  because our goal is to study star formation driven outflows, and AGN are also known to drive outflows at $z\sim2$ (e.g. \citealt{2014ApJ...787...38F}, \citealt{2017arXiv170310255L}).  However, there may be low luminosity AGN that were not detected with these methods.  There is an observational bias against identifying AGNs at all wavelengths in low-mass galaxies (\citealt{2017ApJ...835...27A}).  This bias may lead to some galaxies that host AGN being included in the sample. AGN typically have higher \NII/\Ha \ and \OIII/\Hb \ ratios because of harder ionization coming from the accretion disk, which is consistent with the line ratios we find in the broad component.

With integral field spectroscopy it is possible to create spatially resolved line ratios (\citealt{2014ApJ...781...21N}, \citealt{2010ApJ...711.1291W}), which can be used to determine if a galaxy in the ``composite'' region of the N2-BPT diagram has an AGN. Using spatially-resolved emission line maps, \cite{2014ApJ...781...21N} found some galaxies that lie in the composite region of the BPT diagram host AGN. The cores of these galaxies lie in the AGN region while the outer edges lie in the star-forming region.  The high  \NII/\Ha \ and \OIII/\Hb \ ratios of the core indicated the presence of an AGN that would not be detected in spectra of low spatial resolution. 

It is unlikely that we are detecting outflows driven primarily by AGN because AGN-driven outflows have more extreme kinematics  compared to star-formation driven outflows. AGN driven outflows are typically 500-5000 km/s (\citealt{2014ApJ...796....7G}, \citealt{2014ApJ...787...38F}, \citealt{2017arXiv170310255L}) which is much faster than typical outflow velocities from star-forming galaxies (300-550 km/s) (\citealt{2009ApJ...701..955S}, \citealt{2012ApJ...761...43N}).  The velocity difference between the narrow and broad components in AGN driven outflows is 100-500 km/s (\citealt{2017arXiv170310255L}) while star-formation driven outflows typically have velocity offsets of $<100$ km/s (\citealt{2012ApJ...761...43N}).

Although we have removed all X-ray, IR, and optically identified AGN from this sample, we can not completely rule out some contribution to the emission lines from low mass, low luminosity black holes.  We used an extremely conservative cutoff for optically identified AGN candidates.  If low-luminosity AGN mixing is a significant source of emission in the stacks, then the presence of AGN would have to be extremely widespread among $z\sim2$ galaxies that otherwise appear to be dominated by star formation alone (\citealt{2014arXiv1409.6522C}).  Since starformation rates are higher at $z\sim2$ than at $z\sim0$, weak AGN would not be bright enough to significantly change line ratios (\citealt{2014arXiv1409.6522C}).  It seems much more likely that shocks could be commonplace at $z\sim2$ due to the high SFRs, instead of AGN being ubiquitous.  If there is any AGN contribution it is likely small.

\subsection{Outflows}

One interpretation of the broad component is that it traces ionized outflowing materials (\citealt{1988Natur.334...43B}, \citealt{1990ApJS...74..833H}, \citealt{1992Ap&SS.197...77G}, \citealt{1993AJ....105..486P}, \citealt{1996ApJ...462..651L}, \citealt{2001AJ....121..198V}, \citealt{2004ApJ...602..181C}, \citealt{2007ApJ...671..358W}, \citealt{2008MNRAS.383..864W}, \citealt{2009ApJ...701..955S}, \citealt{2012ApJ...761...43N}, \citealt{2013ApJ...768...75R}, \citealt{2014ApJ...785...75G}, \citealt{2015A&A...583A..99F}, \citealt{2015ApJ...814...83L}).  In this section, we interpret the broad component as a photoionized outflow, calculate the mass loading factor $\eta$ (outflow mass rate/SFR), and compare to other observations as well as simulations.

Using some assumptions about the outflow velocity, radius, temperature, and density we can convert the BFR into an estimate of $\eta$.  We adopt the outflow model from \cite{2011ApJ...733..101G}. This model assumes the broad component to be photoionized and the emission of \Ha \ to be a result of case-B recombination.   

The model assumes a spherical outflow with a constant velocity (cf. \citealt{2010ApJ...717..289S}).  The mass outflow rate, $\dot{M}_{\text{out}}$ can  be calculated as:

\begin{equation}
\dot{M}_{\text{out}} = \frac{1.36 m_H}{\gamma_{\text{H}\alpha}\text{n}_{\text{e}}}\left( L_{H\alpha}  \frac{F_{\text{broad}}}{F_{\text{narrow}}} \right) \frac{V_{\text{out}}}{R_{\text{out}}} 
\end{equation}

where $m_H$ is the atomic mass of hydrogen, $V_{\text{out}}$ is the velocity of the outflow, $R_{\text{out}}$ is the radius,  $\gamma_{\text{H}\alpha}$ is the \Ha \ emissivity, $\text{n}_{\text{e}}$ is the electron density in the outflow, $L_{H\alpha}$ is the total extinction corrected \Ha \ luminosity, and $F_{\text{broad}}/F_{\text{narrow}}$ is the BFR.

We attempt to measure each component of Equation 1 from our stacks (as described below), but sometimes we do not have sufficient signal to do so.  For physical parameters we cannot estimate, we adopt values from \cite{2012ApJ...761...43N} (hereafter N12) who preformed a similar analysis on 27  star-forming galaxies at $z\sim2$.  The sample from N12 have a similar mass range to our sample but have higher SFRs ($\sim90$ \msol/yr on average) which may lead to physical differences.  However, N12 is currently the most similar study to ours with measurements of the parameters in Equation 1, and we use their values when we are unable to measure them from our data.

The electron density for the outflow can be measured using the broad \SIIratio \ ratio (\citealt{1989agna.book.....O}, \citealt{2012ApJ...761...43N}, \citealt{2016ApJ...816...23S}).  We attempt to measure this ratio for the broad components for the stacks.  The flux in the \SII \ lines is low which results in a large measurement uncertainty in the ratio.  We are unable to constrain the density using the broad component from this work.  We adopt the value used by N12 of $50$ $^{+550}_{-50}$ cm$^{-3}$.

The term ${V_{\text{out}}}/{R_{\text{out}}}$ is the inverse of the characteristic timescale of the outflow.  In an attempt to measure the radius of the outflow we made a stack of the 2D spectra and attempted to find broad flux in the spatial direction (e.g. \citealt{2006ApJ...647..222M}, \citealt{2017arXiv170310255L}).  We were unable to measure a spatially extended component in the stacked spectrum. For R$_{\text{out}}$ we adopt the value of 3 kpc as measured by N12.  This value is reasonable given the angular size at this redshift is $\sim8$ kpc arcsec$^{-1}$ and our best seeing is $\sim$0\farcs6.  For $V_{\text{out}}$, we use the  ``maximum" velocity of the outflow defined as $\text{V}_{\text{max}}=|\Delta\text{v}_{\text{B}}$ - $2\sigma_{\text{B}}|$ (\citealt{2011ApJ...733..101G}, \citealt{2015arXiv150700346W}).  This value represents the velocity of the outflow if one assumes the outflow is spherically symmetric with a constant velocity.

We use an \Ha \ emissivity  of $3.56 \times 10^{-25}$ erg cm$^3$ s$^{-1}$ which assumes an electron temperature of  $T_e = 10^4$ K.

If we use the \cite{1998ARA&A..36..189K} relation between SFR and \Ha \ luminosity corrected for a Kroupa IMF ($\text{SFR[\msol   \ yr}^{-1}]=7.9 \times 10^{-42} \ \text{L}_{H\alpha} [\text{ergs s}^{-1}$]), we can divide Equation 1 by SFR and simplify to:

\begin{equation}
\eta   \approx 2.0 \left( \frac{50 \text{ cm}^{-3}}{\text{n}_{\text{e}}}\right) \left( \frac{V_{\text{out}}}{300 \text{ km s}^{-1}}\right) \left( \frac{3\text{ kpc}}{R_{\text{out}}}\right) \left( \frac{F_{\text{broad}}}{F_{\text{narrow}}}\right) 
\label{etaeqn}
\end{equation} 

\begin{figure}
    \centering
    \includegraphics[width=0.48\textwidth]{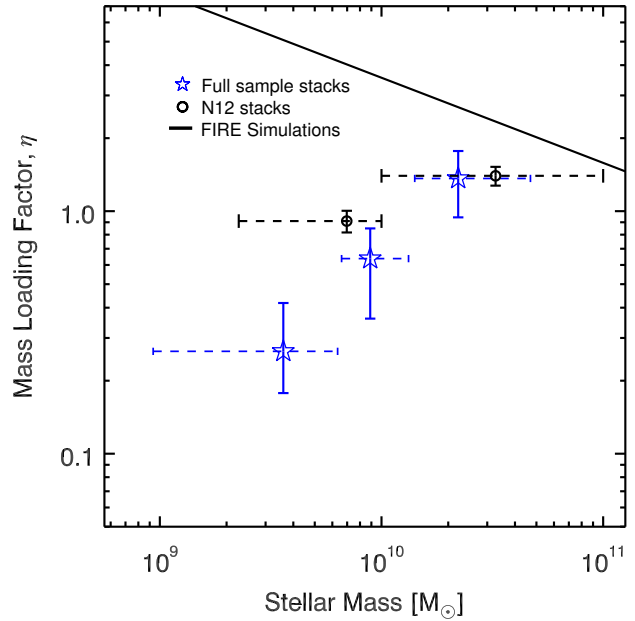}
    \caption{The mass loading factor as a function of mass.  The $z\sim2$ stacks from Figure \ref{quadplot} are shown as blue stars.  Measurements of $\eta$ from N12 are shown as black circles.   We also include the $\eta$ vs. mass relationship found from the FIRE simulations (\citealt{2015arXiv150103155M}).   The conversion from BFR to $\eta$ is  described in detail in Section 5.1.}
    \label{etaplot}
\end{figure}

Figure \ref{etaplot} shows $\eta$ calculated for each stack and the values are listed in Table \ref{tablehalpha}.  The error calculation includes measurement uncertainties from the BFR and the FWHM$_{\text{B}}$.  We do not include errors in the radius, electron density, and temperature assumed, and including these errors would increase the error on the mass loading factor by at least an order of magnitude.  Figure \ref{etaplot} also includes the mass loading factor measured by N12.  The measurements from N12 are higher than those from the MOSDEF stacks despite having similar BFR measurements.  N12 use $V_{\text{out}}=400$ km s$^{-1}$ but we use $\text{V}_{\text{out}}=|\Delta\text{v}_{\text{B}}$ - $2\sigma_{\text{B}}|$ which results in a lower velocity compared to N12 by 50-100 km s$^{-1}$.

We compare our results to the FIRE cosmological galaxy formation simulations with explicit stellar feedback (Hopkins et al. 2014, Muratov et al. 2015).  Interestingly, $\eta$ increases as a function of mass which is contrary to what we expect from the FIRE simulations (\citealt{2015arXiv150103155M}) but are consistent with the results from N12.  This difference is likely explained by our inability to detect low velocity outflows.  The speed of outflows increases as a function of SFR and galaxy stellar mass which is seen in observations (\citealt{2005ApJ...621..227M}, \citealt{2009ApJ...692..187W}) and simulations (\citealt{2015arXiv150103155M}, \citealt{2015arXiv150800007C}).   At low outflow speeds (FWHM$_{\text{B}}<275$ km s$^{-1}$), the emission from the outflow may be indistinguishable from the emission from the HII regions.  To quantify the detectability of low velocity outflows we created two tests using simulated spectra where we can control the BFR and FWHM$_{\text{B}}$ of galaxies used to make stacks.  We can then measure the BFR and FWHM$_{\text{B}}$ of the stacks and check if they are representative of the input galaxy parameters.

In the low velocity test (Test 1), the input FWHM$_{\text{B}}$ increase from 100 to 800 km s$^{-1}$ between stellar masses of $10^{9}$ M$_*$ and $10^{11}$ M$_*$.   For Test 1, galaxies below $10^{10}$ M$_*$ have velocities below 300 km s$^{-1}$.  In the high velocity test (Test 2), the input FWHM$_{\text{B}}$ increase from 275 to 550 km s$^{-1}$ as a function of stellar mass.   Both tests use the same distribution of BFRs and a FWHM$_{\text{N}}$ of 200 km s$^{-1}$.  The slope of the BFR vs. M$_*$ for the tests is the same as the slope of $\eta$ vs. M$_*$ form FIRE simulations (eq. 8 from \cite{2015arXiv150103155M}). We normalize this relation using simplified version of our equation 2, $\eta=\text{C}*$BFR and determine constant C using our highest mass stack (BFR=0.68 at log M$_*$=10.4), where the effects of low velocity outflows should be minimal.  We use the distribution of SFRs and stellar masses of the sample (shown in Figure 1) to make stacks by mass and by SFR.

\begin{figure}
    \centering
    \includegraphics[width=0.49\textwidth]{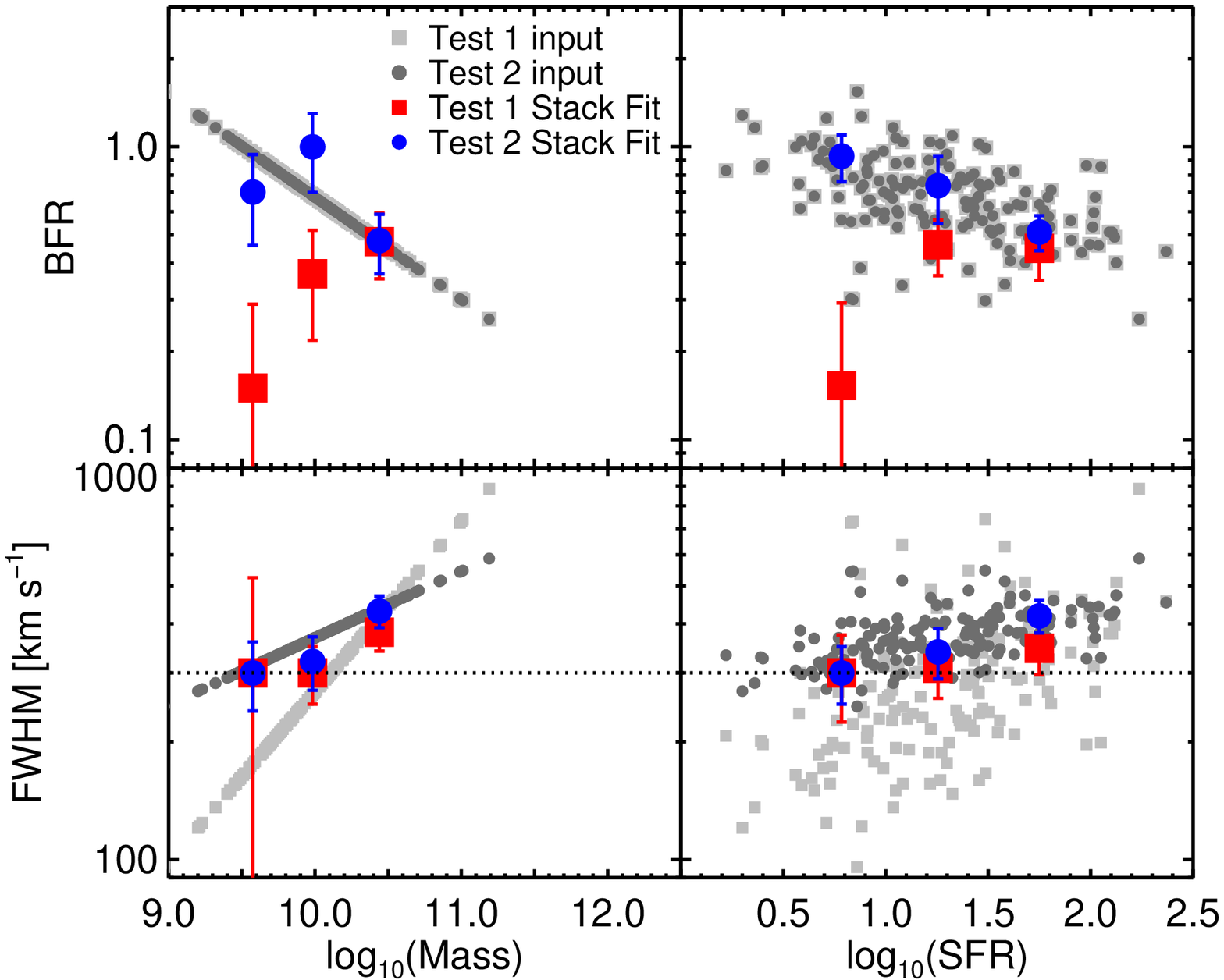}
    \caption{Results from two tests where we added broad components to simulated spectra, created stacks, and fit the stacks using the method described in Section 3.1.  The BFR for both tests are identical, but the FWHM$_{\text{B}}$ for Test 1 ranges from 100-900 km/s and for Test 2 ranges from 260-500 km/s.  The input BFR and FWHM$_{\text{B}}$ are linear with respect to mass.  The left column shows fits to stacks by mass and the right column shows fits to stacks by SFR.  When the FWHM of the broad component is below 275 km/s the BFR is underestimated.}
    \label{gtest_temp}
\end{figure}

The results of these tests are shown in Figure \ref{gtest_temp}.  In Test 1, stacks underestimated the BFR for galaxies in the low and medium mass stacks.  The stacks in Test 1 show an increase in BFR as a function of mass despite the input galaxies having a decrease with mass. The low and medium mass stacks contain 100\% and 55\% galaxies with FWHM$_{\text{B}} < 275$ km s$^{-1}$ respectively.  The FWHM measured for these stacks is too large compared to the input galaxy values although this is expected because we do not allow the FWHM$_{\text{B}}$ parameter to go below 300 km s$^{-1}$ (see Section 3.4).  The stacks in the test show that we could underestimate the extent of broad flux when the FWHM$_{\text{B}} is < 275$ km s$^{-1}$ for a large fraction of galaxies within the stack.  We cannot measure broad flux at low velocities with the measurements made in this work but this is possible with the data in the FIRE simiulation.

\begin{figure}
    \centering
    \includegraphics[width=0.49\textwidth]{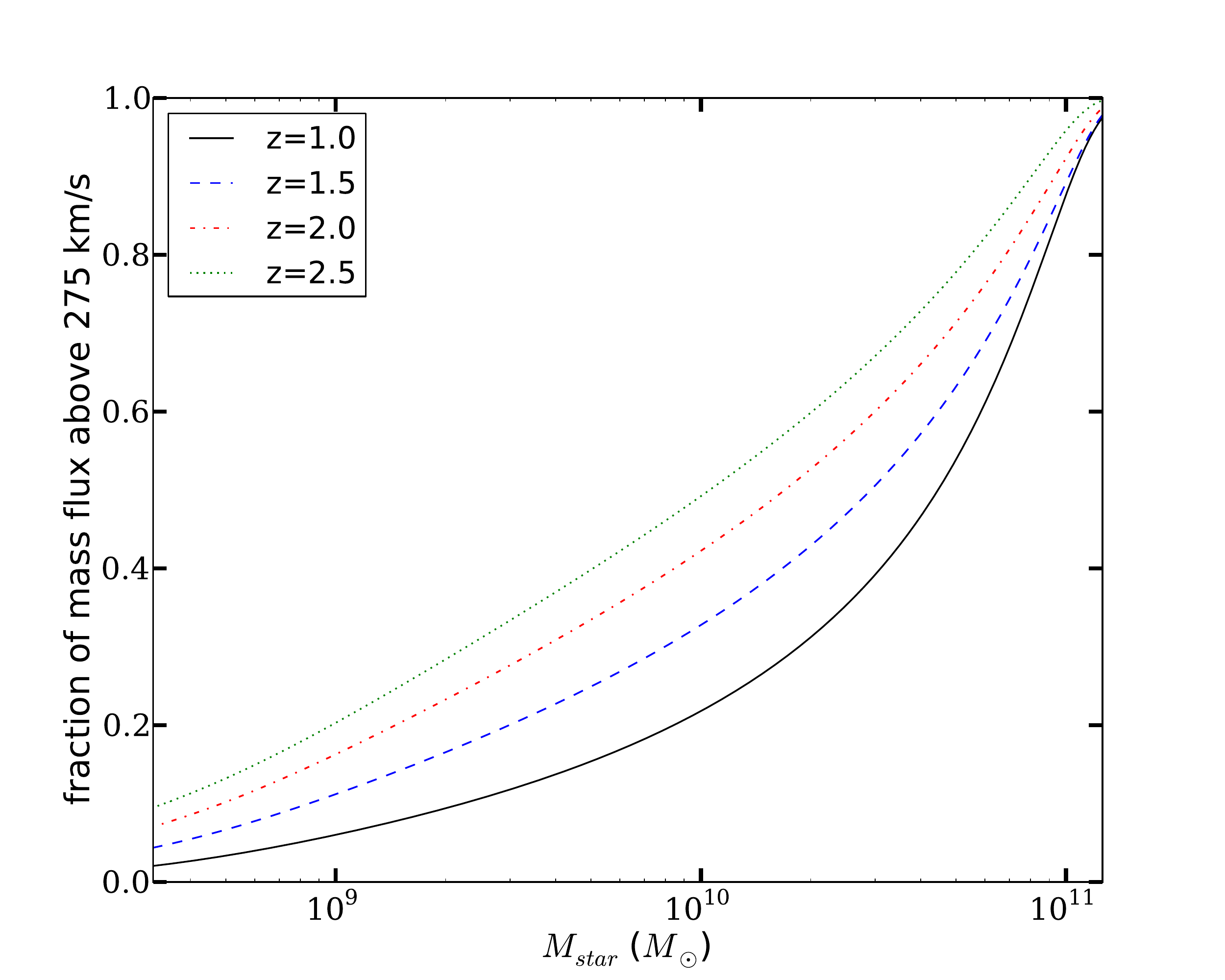}
    \caption{The fraction of mass flux above 275 km/s as a function of stellar mass calculated based on the outflow velocity scaling relations of FIRE simulations from Muratov et al. 2015}
    \label{sasha1}
\end{figure}

In Figure \ref{sasha1} we show the fraction of outflowing material above 275 km s$^{-1}$ versus stellar mass as expected from FIRE simulations. To calculate these fractions we use the fit to median velocities  as a function of halo circular velocity from \cite{2015arXiv150103155M}. We assume log normal velocity distribution at a given circular velocity that matches the 25-75\% velocity distribution range in \cite{2015arXiv150103155M} (their Figure 8). This enables us to calculate the expected fraction of outflows above 275 km s$^{-1}$. We convert the circular velocity to halo mass and then to stellar mass using relations from Behroozi et al. 2013 and show the calculated fraction as a function of M$_*$ for several different redshifts. Direct comparison of our results with FIRE simulation is difficult because they measure outflows at 1/4 of the virial radius and include all possible outflow phases. Nevertheless, if the outflows in the observed high-z galaxies are similar to those in FIRE simulations, it is clear that the high-velocity threshold for broad component will miss the bulk of the outflows in lower mass galaxies. To summarize, the mass loading factor in the low mass galaxy stack in Figure \ref{etaplot} is compatible with the results from the FIRE simulations only if the outflows for low mass galaxies ( $1<10^{10}$ \msol) have low velocities (FWHM$_{\text{B}} <$ 275 km s$^{-1}$).

The mass loading factor for the highest mass stack ($\eta=1.4 _{-0.42}^{+0.41}$) is lower than the predicted value from FIRE at that mass ($\eta=2.6$).  The FWHM$_{\text{B}}$ measured for this stack is 340 $\pm$ 30 km s$^{-1}$ so the difficulty of detecting low velocity outflows should not be a factor (Table \ref{tablehalpha}).  Using a smaller electron density or smaller radius in Equation \ref{etaeqn} would bring these into better alignment however there is no evidence to justify such changes.  An alternate explanation for why $\eta$ in Figure \ref{etaplot} is lower than expected is that some fraction of the outflow is neutral and not visible as a broad \Ha \ component.   Outflows measured in the \Ha \ line are a measure of the ionized component of the outflow, but outflows are multi-phase and have neutral, ionized, and dusty components (\citealt{2015ApJ...814...83L}, \citealt{2015arXiv150700346W}, \citealt{2015A&A...583A..99F}).  Some studies of a small number of local galaxies have measured the neutral phase to have $9-14\times$ as much mass as the ionized phase (\citealt{2015arXiv150700346W}, \citealt{2016MNRAS.tmp..837M}).  Additionally, \cite{2006ApJ...647..222M} measured both Na I absorption and \Ha \ emission in 18 ultra-luminous infrared galaxies and found that there was no correlation between the strength of the Na I absorption and the extended \Ha \ emission.  An undetected neutral component may account for the factor of 2 difference between what is measured here and the FIRE simulations.

The assumptions that the broad component is an outflow and that the broad component is shocked are not mutually exclusive.  The broad component could be a shocked outflow.   If we assume a 100\% collisionally-ionized outflow, the mass loading factor would be smaller by a factor of $\sim2$ (see the appendix of \citealt{2011ApJ...733..101G}).

In conclusion, we estimate the mass loading factor using Equation \ref{etaeqn} (\citealt{2011ApJ...733..101G}) and find generally good agreement with other measurements of $\eta$ at this redshift (\citealt{2012ApJ...761...43N}).  We assumed the electron density and geometry of the outflow were the same as those of other studies (\citealt{2012ApJ...761...43N}) since we were not able to measure these parameters with our sample.  At low masses, $\eta$ increases as a function of mass which is contradictory to the results of the FIRE simulations (\citealt{2015arXiv150103155M}, \citealt{2014MNRAS.445..581H}) where a decrease with mass is predicted.  This difference is best explained by our inability to detect contributions from low velocity (FWHM$<275$ km s$^{-1}$) broad components as shown in Figure \ref{gtest_temp}.  Another feasible explanation is that the outflows have a large neutral component which is not detected because broad \Ha \ emission is sensitive to the ionized component of the outflow.

\section{Implications of Broadened Emission on Estimating Physical Properties of Galaxies}
 
Nebular emission lines provide a means of estimating physical properties of galaxies such as dust extinction, metallicity, electron density, and ionization parameter.  However, most of the calculations assume the emission is coming from photoionized HII regions within the galaxy.  If the broad components we have measured here are a result of shocks, then the inclusion of this flux will affect line ratios and measurements.  In this section, we aim to answer the question: Is it possible that the changes in line ratios between $z\sim0$ and $z\sim2$ are caused by the addition of shocked emission?  

\subsection{The O32, R23, O3N2, and N2 line ratios}
The abbreviations introduced in this section are: 
\begin{displaymath}
\text{N2}=\text{[NII]/H}\alpha
\end{displaymath}
\begin{displaymath}
\text{O32}=\text{[OIII]}\lambda\lambda\text{4959,5007/[OII]}\lambda\lambda\text{3726,3729}
\end{displaymath}
\begin{displaymath}
\text{O3N2}=\text{([OIII]}\lambda\lambda\text{4959,5007)/([NII]/H}\alpha)
\end{displaymath}
\begin{displaymath}
\text{R23}=\text{([OIII]}\lambda\lambda\text{4959,5007 + [OII]}\lambda\lambda\text{3726,3729)/H}\beta\text{)}
\end{displaymath}

\ 

For the MOSDEF sample, \cite{2015ApJ...801...88S} and \cite{2016ApJ...816...23S} showed that the $z\sim2$ galaxies are offset from the $z\sim0$ galaxies in diagnostic diagrams that include  \NII.  Specifically, the galaxies were offset in the N2-BPT, O32 vs. O3N2, and O32 vs. N2 diagrams and did not show any significant offset in the O32 vs. R23 and S2-BPT diagrams.   While there is much speculation, there is no definitive explanation for the offset in diagrams that include \NII \ (eg. \citealt{2014ApJ...795..165S}, \citealt{2016ApJ...826..159S}, \citealt{2014ApJ...785..153M}, \citealt{2016ApJ...828...18M}, \citealt{2015ApJ...801...88S}, \citealt{2016ApJ...816...23S}).  In this section, we test if the offsets in the diagnostic diagrams could be caused only by adding the emission from shocks to the $z\sim0$ spectra.

\begin{figure*}
    \centering
    \includegraphics[width=0.98\textwidth]{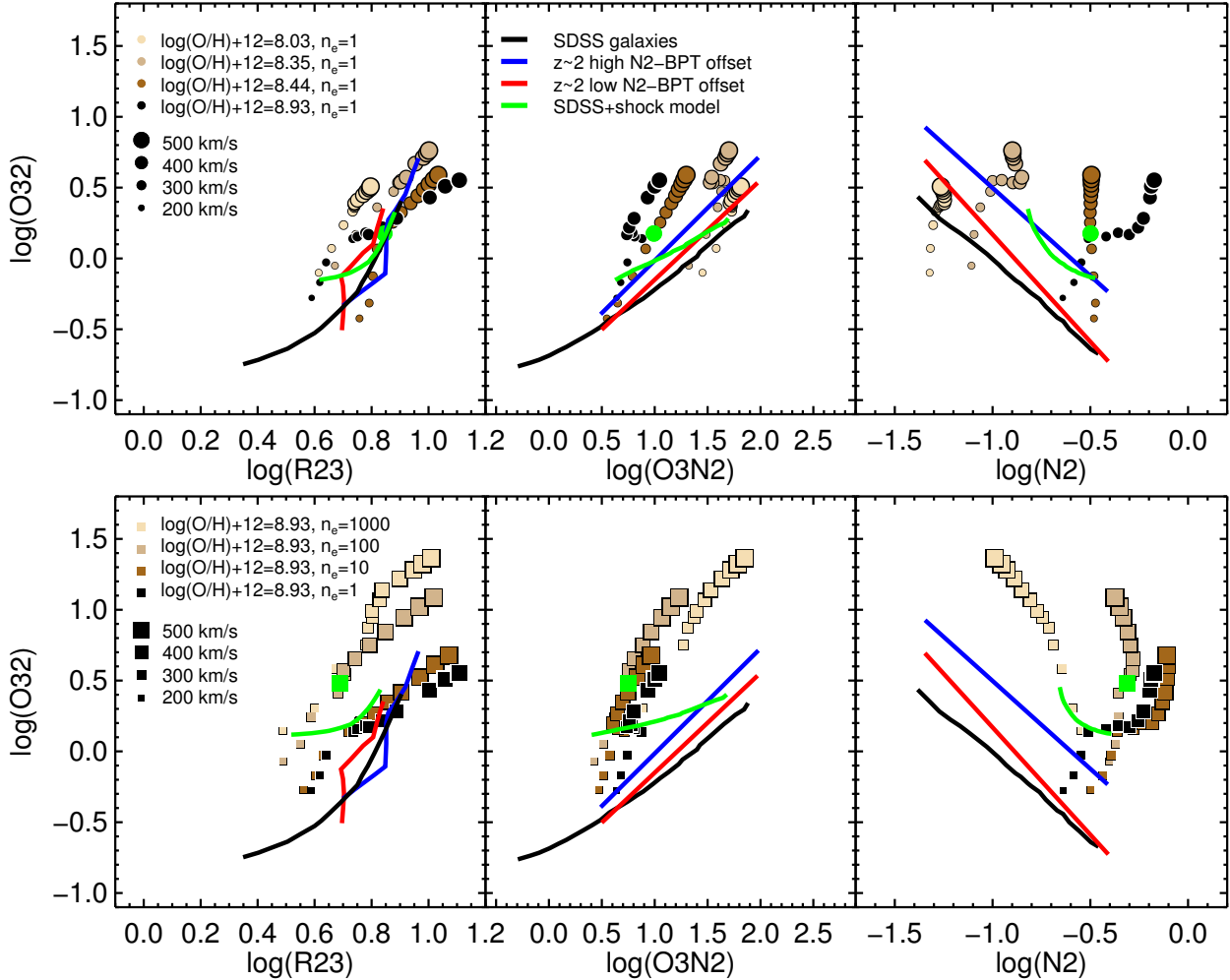}
    \caption{Same as Figure \ref{shocksbpt} but for the O32 vs. R23, O32 vs. O3N2, and O32 vs. N2 diagrams. The tan points and solid black line have the same meaning as in Figure \ref{shocksbpt}.  The blue lines show a running median or linear fit to galaxies in the MOSDEF sample that are more offset than average (compared to SDSS galaxies) in the N2-BPT diagram, and the red lines show a running median or linear fit to galaxies that are below the average MOSDEF galaxy offset from \cite{2016ApJ...816...23S}.  The green point corresponds to the shock model which has (log(O/H)+12 = 8.44, $\text{n}_{\text{e}}=1 \ \text{cm}^{-3}$) in the top row and (log(O/H)+12 = 8.93, $\text{n}_{\text{e}}=100 \ \text{cm}^{-3}$) in the bottom row.  The green line combines the local SDSS data with the green point with 29\% of the flux coming from shocks.}
    \label{shockso32}
\end{figure*}

Figure \ref{shockso32} shows the O32 vs. R23, O32 vs. O3N2, and O32 vs. N2 diagrams with data at $z\sim0$ from SDSS in black and at $z\sim2$ from \cite{2016ApJ...816...23S}.   We overlay the same shock models shown in Figure \ref{shocksbpt}.  The shock models in Figure \ref{shockso32} show the same general trend as the $z\sim2$ data when compared to the SDSS data: no clear offset in O32 vs. R23, a slight offset in O32 vs. O3N2, and a large offset in O32 vs. N2.   

To add the flux from shocks to local data, we select two shock models that are closest to the mean electron density, metallicity, and shock velocity of the MOSDEF sample (290 $^{+88}_{-169}$ cm$^{-3}$ \citealt{2016ApJ...816...23S}, $\text{log(O/H)}+12=8.43$, \citealt{2015ApJ...799..138S}, and shock velocity of 300 km s$^{-1}$).  These models are shown as green points.   We then add the SDSS distribution and shocked data point together assuming a BFR of 0.4 (which is the average BFR for the stacks by \Ha \ and corresponds to 29\% of the total flux being shocked) and plot the result as a green line.  These SDSS+shocks models represent what local galaxies would look like with the addition of the best fitting shock model from Figure \ref{shocksbpt}.

The SDSS+shock data in Figure \ref{shockso32} generally show good agreement with the $z\sim2$ galaxies: no clear offset in O32 vs. R23, a slight offset in O32 vs. O3N2, and a large offset in O32 vs. N2.   The O32 vs. R23 and O32 vs. N2 diagrams for the $\text{n}_{\text{e}}=100 \ \text{cm}^{-3}$ SDSS+shock models (bottom row) are higher than expected but could be explained because these models are at a metallicity of  log(O/H)+12 = 8.93 which is higher than the average $z\sim2$ galaxy metallicity.  Generally, the N2 ratio decreases with decreasing metallicity and the R23 ratio increases with decreasing metallicity.  A decrease in the metallicity of these two models would bring them into better agreement with the $z\sim2$ galaxies.

These models assume a single shock velocity, electron density, and metallicity for the whole sample.  Galaxies at $z\sim2$ have a wide range of metallicities (\citealt{2015ApJ...799..138S}), electron densities (\citealt{2016ApJ...816...23S}), and shock velocities (\citealt{2014ApJ...781...21N}, Table 2).  It is likely that the broad components occupy some region of the N2-BPT diagram based on these physical properties, not just a single point.  We can constrain some parameters by looking at which shock models are unreasonable compared to the $z\sim2$ data in Figure \ref{shockso32}.  At velocities above 400 km/s, the SDSS+shock models would be significantly higher than the $z\sim2$ data in the O32 vs N2 diagram, and at velocities below 250 km/s, the SDSS+shock models would be lower than the $z\sim2$ data in the O32 vs N2 and O32 vs O3N2 diagrams. Changing the metallicity of the shock models (top row) does not result in large shifts in line ratio space except in the O32 vs N2 diagram.  All of the metallicities would fit the data except for log(O/H)+12 = 8.03 which would be too low in the O32 vs N2 diagram.  There is little change with electron density in Figure \ref{shockso32} (bottom row), but there is a large dependence in the S2-BPT diagram (Figure \ref{shocksbpt}).  We can only place limits on the broad \SII/\Ha, and the $3\sigma$ limits are consistent with the lowest electron density models.  Therefore, we conclude that all of the electron densities in the SDSS+shock models would match the $z\sim0$  to $z\sim2$  offsets.

For the O32 vs. R23, O32 vs. O3N2, and O32 vs. N2 diagrams adding shocks to SDSS data at $z\sim0$ could shift the line ratios towards the values measured at  $z\sim2$.  Shock velocities of 250-400 km/s with metallicies ranging from log(O/H)+12 = 8.35-8.93 and electron densities between $\text{n}_{\text{e}}=1-1000 \ \text{cm}^{-3}$ are plausible.  This does not prove that the offset is caused by shocks, only that they are a possibility.  Our results do not rule out contribution from AGN as a driver of these offsets since AGN occupy similar regions of diagnostic line-ratio diagrams as shocks.

\subsection{The N2-BPT diagram}

A great deal of study has been done on the cause of the offset of $z\sim2$ galaxies from the $z\sim0$ galaxies in the  N2-BPT diagram (\citealt{2005ApJ...635.1006S}, \citealt{2006ApJ...644..813E}, \citealt{2008ApJ...678..758L}, \citealt{2016ApJ...828...18M}, \citealt{2014ApJ...795..165S}, \citealt{2015ApJ...801...88S}, \citealt{2017ApJ...836..164S}).  Here, we calculate the same shock+SDSS models for the N2-BPT diagram and calculate where the broad component should lie if it is due to shocks.

\begin{figure}
    \centering
    \includegraphics[width=0.48\textwidth,height=0.48\textwidth]{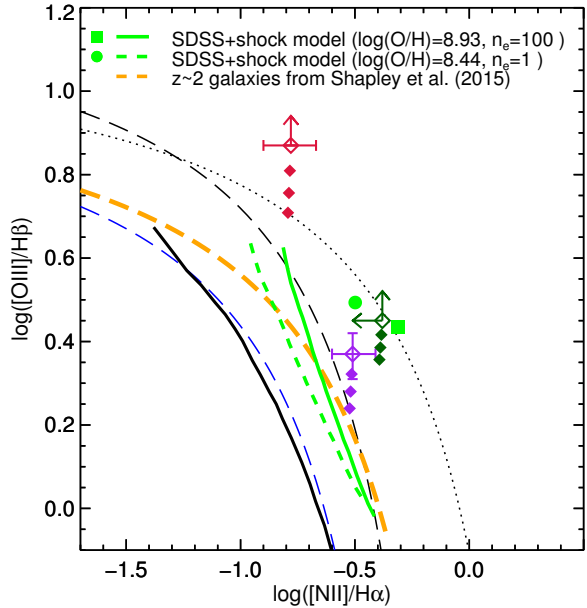}
    \caption{The N2-BPT diagram with the SDSS+shock and broad component predictions.  The blue, orange, and black lines have the same meaning as in Figure \ref{bptdiagram}.  The diamonds are the broad components from Figure \ref{bptdiagram}.  The green lines are the SDSS+shock models.  The dashed green line is the high electron density model and the solid green line is the low metallicity model.  The green circle corresponds to the shock model which has ($v=300$ km s$^{-1}$, log(O/H)+12 = 8.44, $\text{n}_{\text{e}}=1 \ \text{cm}^{-3}$) and the green square corresponds to the shock model which has ($v=300$ km s$^{-1}$, log(O/H)+12 = 8.93, $\text{n}_{\text{e}}=100 \ \text{cm}^{-3}$).  These points are the best match to the metallicity, electron density, and shock velocity for the entire sample.}
    \label{shockspredict}
\end{figure}

Figure \ref{shockspredict} shows the SDSS+shock model along with the location of the $z\sim2$ galaxies from \cite{2015ApJ...801...88S}.  The SDSS+shock model lines show generally good agreement with the line from \cite{2015ApJ...801...88S}.  This implies that adding shocks to the spectra of local galaxies could, in part, explain the offset of the $z\sim2$ galaxies.  This conclusion comes with the caveat that we are unable to completely rule out some contribution from AGN in our stacks.  Since AGN have similar line ratios to shocks the same argument holds that low-luminosity AGN (instead of shocks) could explain the offset of the $z\sim2$ galaxies.

\subsection{The S2-BPT diagram}
As shown in Figure \ref{shocksbpt}, the shocked \SII/\Ha \ ratios that best match the properties of the MOSDEF galaxies are offset to higher values than the SDSS data.  If shocks are the cause of the offset in the N2-BPT diagram, then one might also expect an offset in the S2-BPT diagram as well. However, there is no measured offset between the SDSS and the $z\sim2$ data in the S2-BPT diagram (\citealt{2015ApJ...801...88S}).  We have two possible explainations: the electron density dependence on the shocked \SII/\Ha \ ratio and contribution from diffuse ionized gas.

The electron density of the MOSDEF sample is 290 $^{+88}_{-169}$ cm$^{-3}$, and the shocked line ratios for that particular density would lie between the 100 and 1000 $\text{cm}^{-3}$ shock models.  The SDSS galaxies also lie between the 100 and 1000 $\text{cm}^{-3}$ shock models (see Figure \ref{shocksbpt}). It is possible that the shocked \SII/\Ha \ ratio for and electron density of 290 cm$^{-3}$ lies close to the SDSS distribution.  If this is the case, including the shocked emission would not change the \SII/\Ha \ ratios of the $z\sim0$ galaxies much. The small difference between the shocked \SII/\Ha \ ratio and the photoionized \SII/\Ha \ ratio could explain the lack of an offset in the S2-BPT diagram. 

Another explaination for no offset in the S2-BPT diagram could be because of less contribution from diffuse ionized gas at $z\sim2$ compared to $z\sim0$.  The fraction of \Ha \ emission from diffuse ionized gas decreases as $\Sigma_{H\alpha}$ increases (\citealt{2007ApJ...661..801O}).  As emission from diffuse ionized gas decreases, the \SII/\Ha \ ratio decreases while \OIII/\Hb \ stays the same (\citealt{2017MNRAS.466.3217Z}, \citealt{2017arXiv170804625S}).  Since galaxies at $z\sim2$  have higher SFR (\citealt{2009ApJ...692..778R}, \citealt{1998ApJ...498..106M}, \citealt{2014ARA&A..52..415M}) and are more compact (e.g. \citealt{2005ApJ...630L..17T}, \citealt{2003MNRAS.343..978S}, \citealt{2005ApJ...635..959B}) they have higher $\Sigma_{H\alpha}$ which implies they will have less contribution from diffuse ionized gas if these local relations hold at $z\sim2$.  Local galaxies with high $\Sigma_{SFR}$ do lie at lower \SII/\Ha \  at a given \OIII/\Hb \  than those with low $\Sigma_{SFR}$ on the S2-BPT diagram (\citealt{2016ApJ...828...18M}).  If $z\sim2$ galaxies follow these same trends, then we should expect a lower \SII/\Ha \ ratio at a given \OIII/\Hb \ ratio.  Therefore, the lack of an offset in the S2-BPT diagram could be because less contribution from diffuse ionized gas causes the narrow emission to lie at lower \SII/\Ha \ than average $z\sim0$ galaxies while the broad emission is higher \SII/\Ha \ due to shocks.  In this scenario, these effects compete with each other and ultimately cancel each other out, resulting in no net offset in the S2-BPT diagram.

\subsection{Estimating SFR from \Ha}

The presence of shocks can also affect measurements made from single emission lines such as SFR from \Ha. If the broad emission should be removed when calculating SFR then not doing so would overpredict the SFR.  Given the measured BFRs, SFRs would be overpredicted by 15, 40, and 68\%, respectively in our low, medium, and high stellar mass stacks.   However, given the large number of systematic uncertainties in these calculations (extinction curves, nebular vs. continuum extinction, initial mass functions, and star formation histories), a 15-70\% offset could go undetected. Furthermore, despite only detecting a BFR of 0.15 in the lowest mass bin, the contribution from broad emission may be higher because of our inablility to detect broad emission  with FWHM $<300$ km/s.

\section{Conclusion} 

We present results from the MOSFIRE Deep Evolution Field (MOSDEF) survey on broad emission from the nebular emission lines \Ha, \NII, \OIII, \Hb, and \SII.  After removing known AGN, merging galaxies, and lines affected by skylines, we study broad flux by fitting the emission lines of individual galaxies and stacks using narrow and broad Gaussian components.  The broad flux accounts for 10-70\% of the flux in nebular emission lines when detected.   For individual galaxies, we find no correlations between the BFR as a function of mass, SFR, sSFR, or $\Sigma_{\text{SFR}}$, but there is a strong correlation with higher S/N galaxies and a broad component detection.

We calculate \SII/\Ha, \NII/\Ha, and \OIII/\Hb \ line ratios for the narrow components, broad components, and the single Gaussian fits.  Compared to what one would obtain using a single Gaussian, the broad components have higher  \NII/\Ha \ and \OIII/\Hb \ line ratios.  When placed on the BPT diagram (Figure \ref{bptdiagram}) the broad components for stacks lie within the composite star-forming/AGN region. We compare the locations of the broad component line ratios to shock models from \cite{2008ApJS..178...20A} and conclude that the broad emission could be explained by shocks.  The locations of the broad components could also be explained by contribution from low-luminosity AGN that may have been included in the stack.

We estimate the mass loading factor and we find generally good agreement with other measurements of $\eta$ at this redshift (\citealt{2012ApJ...761...43N}).  At low masses, $\eta$ increases as a function of mass.  This result is contradictory to the results of the FIRE simulations (\citealt{2015arXiv150103155M}, \citealt{2014MNRAS.445..581H}) where a decrease with  mass is predicted.  This difference is best explained by our inability to detect contributions from low velocity (FWHM$<275$ km s$^{-1}$) broad components as shown in Figure \ref{gtest_temp}.  Another feasible explanation is that the outflows have a large neutral component which is not detected because broad \Ha \ emission is sensitive to the ionized component of the outflow. 

 We combine the shock models from \cite{2008ApJS..178...20A} with local line ratios from SDSS and calculate where these galaxies would lie on several emission line diagnostic diagrams.  We compare these SDSS+shock models to the emission line properties of $z\sim2$ galaxies to test if only the addition of shocks could account for the shifts in emission line diagrams between $z\sim0$ and $z\sim2$.   If we add a 29\% shocked component to SDSS data at $z\sim0$, the  N2-BPT, O32 vs. O3N2, and O32 vs. N2  diagrams have similar offset line ratios to the observed $z\sim2$ data from \cite{2016ApJ...816...23S} and \cite{2015ApJ...801...88S}.  There is no offset in the O32 vs. R23 diagram which is also seen in \cite{2016ApJ...816...23S} at $z\sim2$.  The lack of an offset in the S2-BPT diagram seen at $z\sim2$ may be due to the strong dependence of shocked emission on electron density or from decreased contribution from diffuse ionized gas.  Since AGN have similar ratios to shocked emission, it is also possible that AGN contribution could explain the positions of $z\sim2$ galaxies instead of shocks.

If the offsets between $z\sim0$ and $z\sim2$ galaxies in emission line diagnostic diagrams are caused by outflowing, shocked gas, then the contribution from the shocks can be subtracted to isolate emission from HII regions when calculating star formation rate. Given the measured BFRs, SFRs would be overpredicted by 15, 40, and 68\%, respectively in our three mass bins.   However, given the large number of systematic uncertainties in these calculations (extinction curves, nebular vs. continuum extinction, initial mass functions, and star formation histories), a 15-70\% offset could go undetected.

In this work, we have shown that galaxies at $z\sim1-3$ have a broad component and that the origin of this emission is likely shocks or outflows.  In either case, the broad emission may complicate how we interpret galaxy properties measured from emission lines (\citealt{2013ApJ...774..100K}, \citealt{2014ApJ...781...21N}).  Additionally, a better estimate of the electron density of the outflows would be beneficial because the uncertainty in this measurement dominates our error in calculating the mass loading factor.  Future studies of outflows at $z\sim2$ would greatly benefit from increased spatial information.  High spatial resolution integral field unit maps aided by adaptive optics (e.g., with Keck/OSIRIS) will enable us to disentangle if the broad component is from AGN or is truly due to outflows (\citealt{2014ApJ...781...21N}).

\section*{Acknowledgments}

We thank Bili Dong for commenting on an early version of this work and discussing details about current and future work with the FIRE simulation.  We thank the MOSFIRE instrument team for building this powerful instrument.

This work would not have been possible without the 3D-HST collaboration, who provided us the spectroscopic and photometric catalogs used to select our targets and to derive stellar population parameters.  We are grateful to I. McLean, K. Kulas, and G. Mace for taking observations for us in May and June 2013.  We acknowledge support from an NSF AAG collaborative grant AST-1312780, 1312547, 1312764, and 1313171, and archival grant AR-13907, provided by NASA through a grant from the Space Telescope Science Institute.

This work is also based on observations made with the NASA/ESA Hubble Space Telescope (programs 12177, 12328, 12060-12064, 12440-12445, 13056), which is operated by the Association of Universities for Research in Astronomy, Inc., under NASA contract NAS 5-26555.

NAR is supported by an Alfred P. Sloan Research Fellowship.

The data presented in this paper were obtained at the W.M. Keck Observatory, which is operated as a scientific partnership among the California Institute of Technology, the University of California and the National Aeronautics and Space Administration. 
The Observatory was made possible by the generous financial support of the W.M. Keck Foundation. The authors wish to recognize and acknowledge the very significant cultural role and reverence that the summit of Mauna Kea has always had within the indigenous Hawaiian community.  We are most fortunate to have the opportunity to conduct observations from this mountain.

\def\arraystretch{1.5}

\begin{deluxetable*}{lllllllll}
\tablecolumns{9}
\tablecaption{Measurements from the Stack Fits}
\tablehead{ 
\colhead{Parameter\tablenotemark{a}} &
\colhead{Avg.\tablenotemark{b}} & 
\colhead{Range\tablenotemark{c} } & 
\colhead{BFR\tablenotemark{d}}  & 
\colhead{BFR$_{\text{min}}$\tablenotemark{e}}  & 
\colhead{BFR$_{\text{max}}$\tablenotemark{f}}  & 
\colhead{FWHM$_{\text{B}}$\tablenotemark{g}}  & 
\colhead{$\Delta\text{V}_{\text{B}}$\tablenotemark{h}} & 
\colhead{$\eta$\tablenotemark{i}}  
} 
\startdata
(\Ha)-$z\sim2$ & 9.56 & 8.97 | 9.80 & 0.15 $   {+0.071 \atop{-0.041}} $   & 0.056 & 0.35 &300 $\pm$ 200 & -9.0 $\pm$ 20 & 0.26 $ {+0.15 \atop{ -0.086}}$  \\ 
(\Ha)-$z\sim2$ & 9.95 & 9.82 | 10.1 & 0.40 $   {+0.092 \atop{-0.15}} $   & 0.15 & 0.70 &300 $\pm$ 60 & 16. $\pm$ 10 & 0.64 $ {+0.21 \atop{ -0.28}}$  \\ 
(\Ha)-$z\sim2$ & 10.3 & 10.1 | 10.7 & 0.68 $   {+0.15 \atop{-0.18}} $   & 0.35 & 1.2 &340 $\pm$ 30 & -10. $\pm$ 6 & 1.4 $ {+0.41 \atop{ -0.42}}$  \\ 
 (\OIII)-$z\sim2$ & 9.53 & 8.97 | 9.80 & 0.11 $\pm 0.028 $   & 0.029 & 0.20 &390 $\pm$ 100 & -63 $\pm$ 40 & - \\ 
 (\OIII)-$z\sim2$ & 9.92 & 9.82 | 10.1 & 0.56 $\pm 0.12 $   & 0.21 & 0.92 &300 $\pm$ 20 & -18 $\pm$ 8 & - \\ 
 (\OIII)-$z\sim2$ & 10.4 & 10.1 | 10.7 & 0.58 $\pm 0.17 $   & 0.060 & 1.1 &300 $\pm$ 30 & 8.5 $\pm$ 9 & - \\ 
 (\OIII)-$z\sim3$ & 9.5 & 9.2 | 9.8 & 0.75 $\pm 0.14 $   & 0.32 & 1.2 &340 $\pm$ 30 & -13 $\pm$ 9  & -  \\ 
 (\OIII)-$z\sim3$ & 10.0 & 9.82 | 10.1 & 0.37 $\pm 0.16 $   & 0.08 & 0.85 &300 $\pm$ 100 & 27 $\pm$ 20  & -  \\ 
 (\OIII)-$z\sim3$ & 10.5 & 10.1 | 11.0 & 0.71 $\pm 0.37 $   & 0.1 & 1.8 &320 $\pm$ 80 & 25 $\pm$ 20  & -  
\tablenotetext{a}{Parameter by which the stack was created.} 
\tablenotetext{b}{The geometric mean of the galaxies included.}
\tablenotetext{c}{Mass range of galaxies included.}
\tablenotetext{d}{Broad Flux Fraction of the stack and 1$\sigma$ errors.}
\tablenotetext{e}{The 3$\sigma$ lower limit on the BFR.}
\tablenotetext{f}{The 3$\sigma$ upper limit on the BFR.}
\tablenotetext{g}{The FWHM of the broad component (km s$^{-1}$) and the 1$\sigma$ error.}
\tablenotetext{h}{The velocity offset between the peaks of the broad and narrow components (km s$^{-1}$). A negative value indicates a blueshift. The 1$\sigma$ error in the velocity offset is included.}
\tablenotetext{g}{The mass loading factor (see Section 5.1).}

    \enddata
\label{tablehalpha}
  \end{deluxetable*}

\def\arraystretch{1.0}

\appendix

\section{One-Dimensional Extraction Software: \texttt{BMEP}}
The MOSDEF team has written software to handle the 2D and 1D reduction process.  The 2D code is described in \cite{2015ApJS..218...15K} and the 1D extraction code, \texttt{BMEP}\footnote{Source code and installation instructions available at: \url{https://github.com/billfreeman44/bmep}}, is described here.  In general, \texttt{BMEP} can be used to extract spectra from any rectified 2D  spectroscopic data including the MOSFIRE Data Reduction Pipeline.

The 2D reduction code outputs 2D spectra that are combined, flat-corrected, cleaned of cosmic rays, and rectified.  We have designed a 1D extraction program that can optimally extract spectra, help the user find the primary object, create a redshift catalog, and ``blindly" extract spectra where there were no obvious emission lines or continuum.

\subsection{Using \texttt{BMEP}}

Reduced 2D spectra have two dimensions: spatial and wavelength.  The overall goal of \texttt{BMEP} is to optimally extract 1D integrated spectra which sums the flux over the spatial dimension and leaves the user with flux vs. wavelength.  The first step in extracting spectra is to find the primary object.  \texttt{BMEP} draws a line over the object's expected position which helps distinguish the primary object from serendipetous objects.  Next, the user interactively bins the data in the wavelength direction to create a flux profile.  Finally, the user fits a Gaussian to the profile.  The center and width of this Gaussian determine the spatial region which to sum as well as the weighting profile for an optimal extraction (\citealt{1986PASP...98..609H}).  In some objects with high S/N the profile is non-Gaussian, and the user can choose to weight by the actual profile instead of the Gaussian fit.

Although the above process sounds simple, it is difficult to determine which wavelengths to bin to create the highest S/N spatial profile.  Many galaxies at high redshift have bright emission lines compared to their continuum (\citealt{2013MNRAS.436.1040S}).  Summing all wavelengths results in a spatial profile dominated by noise which makes finding the center and width of the object impossible.  The benefit of using \texttt{BMEP} is the ability to interactively create the best profile by adding or removing wavelengths when creating the spatial profile.  Additionally, some galaxies do have continuum but summing all wavelengths results in a noisy spatial profile because skylines would be included.  The user can enable a ``continuum mode" which does not include skylines in the spatial profile by removing wavelengths where the variance is higher than the median variance.

After extraction, the spectrum is plotted and can be inspected.  The locations where the user clicked are also drawn in red on the plot.  In noisy spectra, this allows the user to easily find emission lines in the 1D spectra.  Once an emission line is found, the user can fit the line to a Gaussian, input which line it is, and calculate a redshift.  All emission lines and calculated redshifts are saved in a catalog.  A separate program consolidates all the lines fit for each object and calculates the most likely redshift.

In cases where an object had no obvious emission lines or continuum in the 2D spectrum, a ``blind" extraction was performed. For objects with no signal in any bands, the blind extraction uses the expected position of the object calculated from the mask file and uses the same extraction width as the star's width in each filter.    For objects with signal in one or more bands, the blind extraction used the average extraction widths and centers from filters where there was signal from the object.  The widths from each filter are corrected for seeing differences.  These blind spectra allow us to put upper limits on emission lines for spectra if we know the redshift.  Currently, \texttt{BMEP} is only able to read in MOSFIRE mask files for the blind extraction and would need to be modified to be able to read in mask files of a different format.

\subsection{Sub-pixel Extraction Equations}

The optimal extraction used in \texttt{BMEP} is based on \cite{1986PASP...98..609H}.  While testing the software, we compared extractions of a bright object done by several different users.  Some extractions differed in extracted flux by 2-4\% at all wavelengths for some users.  We traced the cause of this to rounding differences between two extraction profiles.  The extraction width is determined by a Gaussian fit to the profile and in some cases, the extracted widths would be 1 pixel different because we rounded the extraction to the nearest whole pixel.   The optimal extraction of \cite{1986PASP...98..609H} sums over an integer number of pixels in the spatial direction to calculate the flux at each wavelength. To fix this, we created a sub-pixel optimal extraction algorithm.  This algorithm extracts the central pixels exactly the same as in \cite{1986PASP...98..609H} but adds a fraction of a pixel at each end.

We base the sub-pixel algorithm on Equation 8 from \cite{1986PASP...98..609H}.  However, it is simplified for MOSFIRE reduction because there is no sky subtraction or cosmic ray rejection needed as these are done during the 2D reduction.  The equations from \cite{1986PASP...98..609H} with these simplifications are: 

\begin{displaymath}
x_b'=\mathbf{R}(c - w)
\end{displaymath}

\begin{displaymath}
x_t'=\mathbf{R}(c + w)
\end{displaymath}

\begin{equation}
\sum_{x_b'}^{x_t'}D = F'_{\text{box}}
\end{equation}

\begin{equation}
\sum_{x_b'}^{x_t'}V = V'_{\text{box}}
\end{equation}

\begin{equation}
\frac{\sum_{x_b'}^{x_t'}PD/V}{\sum_{x_b'}^{x_t'}P^2/V} = F'_{\text{opt}}
\end{equation}

\begin{equation}
\frac{\sum_{x_b'}^{x_t'}P}{\sum_{x_b'}^{x_t'}P^2/V} =  V'_{\text{opt}}
\end{equation}

Unnumbered equations are defining relationships or variables.  Bold letters indicate functions.  $\mathbf{R}$ is the round function, $c$ is the center of the object from the Gaussian fit to the profile, $w$ is half the width to extract, $D$ is the the flux in one pixel, $V$ is the variance in one pixel, $P$ is the weighting profile, $x_b$ is the pixel at the bottom of the profile, and $x_t$ is the pixel at the top of the profile.  The weighting profile comes from the Gaussian fit to the spatial profile.   $F'_{\text{box}}$ is the boxcar flux,  $V'_{\text{box}}$ is the boxcar variance, $F'_{\text{opt}}$ is the optimal flux, and $V'_{\text{opt}}$ is the optimal variance for the \cite{1986PASP...98..609H} algorithm that does not include sub-pixel corrections.  We remove the wavelength subscript for simplification.

We extend this equation to extract a fraction of each pixel.  The central region of extraction is extracted the same as in \cite{1986PASP...98..609H}, then a fraction of the outer pixels are added to this flux.  First we calculate the range which the flux is extracted in the same manner as equations A1-A4.  This is between $xb'$ and $xt'$ which are calculated as follows:

\begin{displaymath}
x_b=\mathbf{L}(c - w +1)
\end{displaymath}

\begin{displaymath}
x_t=\mathbf{L}(c + w)
\end{displaymath}

$\mathbf{L}$ is the ``Floor'' function.  Next, calculate the ``remainder" from the bottom ($R_b$) and the top ($R_t$):

\begin{displaymath}
R_b =1 - ( x_b-x_b' )
\end{displaymath}

\begin{displaymath}
R_t = x_t-x_t'
\end{displaymath}

Now we calculate the weighting for sub-pixel extraction on the edges:
\begin{displaymath}
P_t = P(xt+1)R_t
\end{displaymath}

\begin{displaymath}
P_b = P(xb-1)R_b
\end{displaymath}

The boxcar extraction for the sub-pixel algorithm is:
\begin{displaymath}
B=D(xb-1)R_b
\end{displaymath}

\begin{displaymath}
T=D(xt+1)R_b
\end{displaymath}

\begin{equation}
\sum_{x_b}^{x_t}D + B + T = F_{\text{box}}
\end{equation}

\begin{equation}
\sum_{x_b}^{x_t}V + V_bR_b +V_tR_t = V_{\text{box}}
\end{equation}

Where B and T are the flux from the bottom and top pixels to be added to the central region.  For the optimal extraction this extra flux is:

\begin{displaymath}
B=\frac{P_bD(x_b-1)R_b}{V(x_b-1)}
\end{displaymath}

\begin{displaymath}
T=\frac{P_tD(x_t+1)R_t}{V(x_t+1))}
\end{displaymath}

Calculate sub-pixel flux and variance:

\begin{equation}
\frac{(\sum_{x_b}^{x_t}PD/V) + B + T}{\sum_{x_b}^{x_t}P^2/V+ P_b^2/V_b+ P_t^2/V_t } = F_{\text{opt}}
\end{equation}

\begin{equation}
\frac{(\sum_{x_b}^{x_t}P) + P_b +P_t}{\sum_{x_b}^{x_t}P^2/V+ P_b^2/V_b+ P_t^2/V_t} = V_{\text{opt}}
\end{equation}

If the spatial range to extract are integers, then $R_b=1$, $R_t=0$, $x_b'=x_b+1$, and $x_t'=x_t$, and one can recover the original equations from \cite{1986PASP...98..609H}.

\begin{figure}[h]
    \centering
    \includegraphics[width=0.48\textwidth]{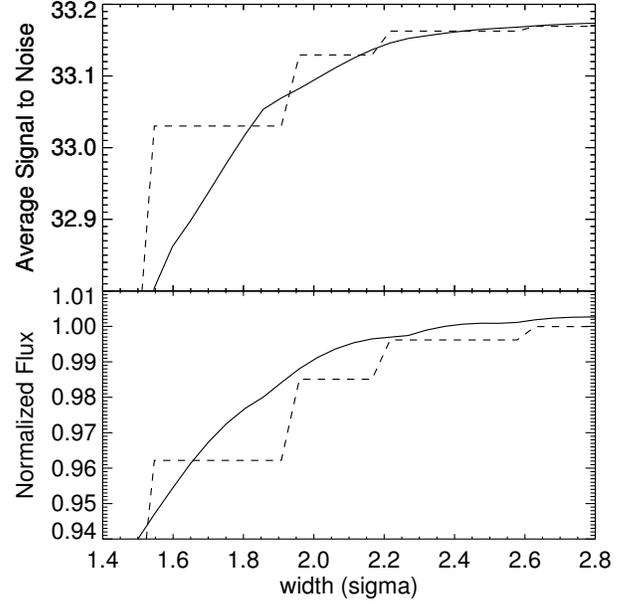}
    \caption{A comparison between sub-pixel and the \cite{1986PASP...98..609H} extractions. The solid line is the sub-pixel extraction and the dashed is the \cite{1986PASP...98..609H} extraction.  This plot is made by first extracting a star normally, then looking for a section of the spectrum that is featureless showing no absorption features, emission lines, or sky lines.  Next, the spectrum is extracted using widths between 1.5 and 5.0 pixels in increments of 0.1 pixels.  The points are plotted as the lines in the figure above.  Because each star has a slightly different width, we convert the width in pixels to a width in ``sigma''.}
    \label{starprofile}
\end{figure}

Figure \ref{starprofile} shows a comparison of the sub-pixel optimal vs. normal optimal.  This figure was produced by selecting a flat, featureless section of a star that has no sky lines.  Within this region, we calculated the average flux and S/N, then we varied the extracted width.  As one might expect, the Horne extraction has steps where the width rounds to the next pixel and the sub-pixel extraction is smooth.  Though the jumps in flux are severe when the width is small, the steps flatten out as width increases.  At the width where we extract (2x FWHM), the jumps between steps is quite small, at worst around 4\%.  However, since we use a standard star to calculate the absolute flux this can cause every flux for a mask to be 4\% different when two different people extract a mask only due to rounding.

\section{Results of Non-Parametric Ratio Estimation}

\begin{figure}
    \centering
    \includegraphics[width=0.49\textwidth]{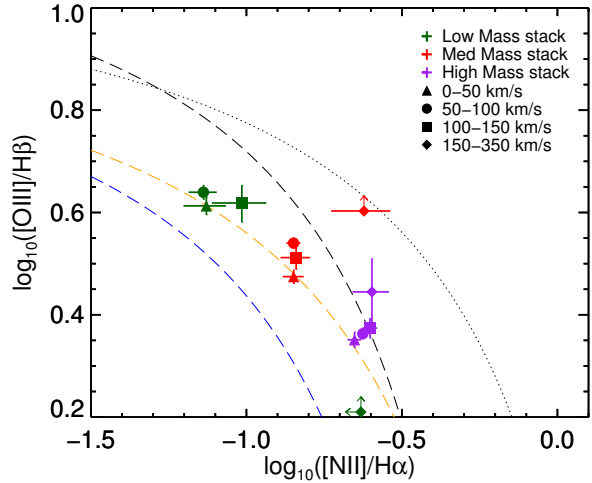}
    \caption{BPT diagram created by integrating the flux over different velocities.}
    \label{numerical_bpt}
\end{figure}

When fitting the broad flux, we assumed the broad flux is Gaussian in shape.    To measure broad emission regardless of shape, we calculate line ratios using the flux at different velocities from the centroid of each line for stacks from Figure \ref{bptdiagram}.  The results are shown in Figure \ref{numerical_bpt}.  This provides a non-parametric measurement of the line ratios as a function of velocity.  The high velocity points are typically higher than the \cite{2003MNRAS.346.1055K} line which is a similar trend to the broad component ratios in Figure \ref{bptdiagram}.  Since the non-parametric measurements have similar results to the fits, it is reasonable to use the fits in our analysis.


\bibliographystyle{apj}
\bibliography{outflows_cites}

\end{document}